\documentclass[sigconf]{acmart}
\makeatletter
\newcommand{\removelatexerror}{\let\@latex@error\@gobble}
\makeatother

\newif\iflong
\longtrue 
\newif\ifdebug
\debugfalse 

\usepackage[section]{placeins}
\usepackage{amsmath}
\usepackage{amssymb}
\usepackage{amsthm}
\usepackage{color}
\usepackage{graphicx}
\usepackage{listings}
\usepackage{mathtools}
\usepackage{stmaryrd}
\usepackage{wrapfig}
\usepackage{balance}
\usepackage[normalem]{ulem}
\usepackage{nccmath}

\usepackage{paralist}
\usepackage[textwidth=1.5cm]{todonotes}
\usepackage[T1]{fontenc}
\usepackage{framed}
\usepackage{enumerate}
\usepackage{enumitem}
\usepackage[all]{xy}
\usepackage{todonotes}
\usepackage{cleveref} 

\usepackage{xspace}
\xspaceaddexceptions{(,\$}

\usepackage{caption}
\usepackage{tikz}
\usetikzlibrary{arrows}
\newcommand*\circled[1]{\tikz[baseline=(char.base)]{
            \node[shape=circle,fill, color=black,draw,inner sep=.5pt] (char) {\small
              {\bf \textcolor{white}{#1}}};}}

\usepackage[vlined,linesnumbered,noresetcount]{algorithm2e}
\SetKwBlock{SubAlgoBlock}{}{end}
\newcommand{\SubAlgo}[2]{#1 \SubAlgoBlock{#2}}
\let\oldnl\nl
\newcommand{\nonl}{\renewcommand{\nl}{\let\nl\oldnl}}

\newtheorem{liveness}{Assumption}

\newboolean{showcomments}
\setboolean{showcomments}{true}
\ifthenelse{\boolean{showcomments}}
{ \newcommand{\mynote}[1]{
	       {\color{blue}{#1}}}}
{ \newcommand{\mynote}[1]{}}
\newcommand{\newtext}[1]{\mynote{#1}}

\SetKw{WhenReceived}{when received}
\SetKw{WhenDelivered}{when received-rdma}
\SetKw{KwTo}{to}
\SetKw{Send}{send}
\SetKw{Broadcast}{broadcast}
\SetKw{KwFrom}{from}
\SetKw{Fun}{function}

\newcommand{\DECISION}{{\tt DECISION}}
\newcommand{\ACCEPT}{{\tt ACCEPT}}
\newcommand{\ACCEPTACK}{{\tt ACCEPT\_ACK}}

\newcommand{\PREPARE}{{\tt PREPARE}}
\newcommand{\PREPAREACK}{{\tt PREPARE\_ACK}}

\newcommand{\PROBE}{{\tt PROBE}}
\newcommand{\PROBEACK}{{\tt PROBE\_ACK}}

\newcommand{\NEWCONFIG}{{\tt NEW\_CONFIG}}

\newcommand{\REPLICATELOGS}{{\tt NEW\_STATE}}

\newcommand{\CONFIGCHANGE}{{\tt CONFIG\_CHANGE}}

\newcommand{\status}{{\sf status}}
\newcommand{\reconstatus}{{\sf rec\_status}}
\newcommand{\probing}{{\sf probing}}

\newcommand{\txn}{{\sf txn}}
\newcommand{\nextv}{{\sf next}}
\newcommand{\phase}{{\sf phase}}
\newcommand{\payload}{{\sf payload}}
\newcommand{\decision}{{\sf dec}}
\newcommand{\vote}{{\sf vote}}
\newcommand{\ballot}{{\sf epoch}}
\newcommand{\cballot}{{\sf new\_epoch}}

\newcommand{\vcballot}{\mathit{cur\_epoch}}
\newcommand{\vballot}{\mathit{e}}

\newcommand{\vphase}{\mathit{phase}}
\newcommand{\vpayload}{\mathit{payload}}
\newcommand{\vtxn}{\mathit{txn}}
\newcommand{\vdecision}{\mathit{dec}}
\newcommand{\vvote}{\mathit{vote}}

\newcommand{\shards}{{\sf shards}}

\newcommand{\client}{{\sf client}}

\newcommand{\leader}{{\sf leader}}
\newcommand{\pos}{\mathit{pos}}
\newcommand{\pload}{\mathit{pload}}

\newcommand{\length}{{\sf length}}

\newcommand{\M}{{\sf members}}

\newcommand{\vm}{\mathit{M}}

\newcommand{\zk}{\text{CS}}
\newcommand{\recovery}{{\sf initialized}}
\newcommand{\vrecovery}{\mathit{initialized}}
\newcommand{\pballot}{{\sf probed\_epoch}}
\newcommand{\rballot}{{\sf recon\_epoch}}
\newcommand{\rshard}{{\sf recon\_shard}}
\newcommand{\pmembers}{{\sf probed\_members}}
\newcommand{\rmembers}{{\sf recon\_members}}
\newcommand{\rleaders}{{\sf recon\_leaders}}

\newcommand{\vpload}{l}

\newcommand{\fploadprocess}{{\sf topload}}

\newcommand{\lltcsone}{{\sf TCS-LL}}

\newcommand{\prefix}[2]{#1\mathpunct{\downharpoonleft}_{#2}}

\newcommand{\LEADER}{\textsc{leader}}
\newcommand{\FOLLOWER}{\textsc{follower}}
\newcommand{\RECOVERING}{\textsc{reconfiguring}}

\newcommand{\START}{\textsc{start}}

\newcommand{\ACCEPTED}{\textsc{prepared}}
\newcommand{\DECIDED}{\textsc{decided}}

\newcommand{\COMMIT}{\textsc{commit}}
\newcommand{\ABORT}{\textsc{abort}}

\newcommand{\TRUE}{\textsc{true}}
\newcommand{\FALSE}{\textsc{false}}
\newcommand{\READYSTATUS}{\textsc{ready}}
\newcommand{\PROBING}{\textsc{probing}}
\newcommand{\INSTALLING}{\textsc{installing}}

\newcommand{\D}{\mathcal{D}}

\newcommand{\Proc}{\mathcal{P}}
\newcommand{\Txn}{\mathcal{T}}

\newcommand{\Shard}{\mathcal{S}}

\newcommand{\Shards}{\mathcal{S}}
\newcommand{\Payload}{\mathcal{L}}
\makeatletter
\renewcommand\subparagraph{\@startsection{subparagraph}{5}{\z@}%
                                       {2ex \@plus1ex \@minus .2ex}%
                                       {-1em}%
                                      {\normalsize\bfseries}}
\makeatother
\newcommand{\Certify}{\ensuremath{{\tt certify}}}
\newcommand{\Decide}{\ensuremath{{\tt decide}}}

\newcommand{\act}{\ensuremath{{\sf act}}}

\newcommand{\committed}{\ensuremath{{\sf committed}}}

\newcommand{\hist}{h}
\newcommand{\cf}{f}
\newcommand{\cg}{g}

\newcommand{\Val}{\ensuremath{{\sf Val}}}
\newcommand{\Obj}{\ensuremath{{\sf Obj}}}
\newcommand{\Ver}{\ensuremath{{\sf Ver}}}

\newcommand{\dec}{\ensuremath{{\sf dec}}}

\newcommand{\tr}[1]{\iflong{}\S\ref{#1}\else{}\cite{ext}\fi}

\newcommand{\rt}{\prec_{\text{rt}}}
\newcommand{\decrel}{\prec_{\text{dec}}}

\newcommand{\ms}{\\[2pt]}

\newcommand{\external}{configuration service\xspace}

\newcommand{\CONFIGPREPARE}{{\tt CONFIG\_PREPARE}}
\newcommand{\CONFIGPREPAREACK}{{\tt CONFIG\_PREPARE\_ACK}}

\newcommand{\OPENCONNECTION}{{\tt CONNECT}}
\newcommand{\OPENCONNECTIONACK}{{\tt CONNECT\_ACK}}
\newcommand{\openconnections}{{\sf connections}}

\newcommand{\emptybuffers}{{\bf flush}\xspace}
\newcommand{\openconnection}{{\bf open}\xspace}
\newcommand{\closeconnection}{{\bf close}\xspace}

\newcommand{\sendrdma}{{\bf send-rdma}\xspace}
\newcommand{\deliverrdma}{{\bf deliver-rdma}\xspace}
\newcommand{\ackrdma}{{\bf ack-rdma}\xspace}
\newcommand{\multiclose}{\xspace{\tt multiclose}\xspace}

\newcommand{\pl}{\mathit{l}}
\newcommand{\setpayload}{\mathit{L}}

\def\BibTeX{{\rm B\kern-.05em{\sc i\kern-.025em b}\kern-.08emT\kern-.1667em\lower.7ex\hbox{E}\kern-.125emX}}

\copyrightyear{2019} 
\acmYear{2019} 
\setcopyright{acmlicensed}
\acmConference[PODC '19]{2019 ACM Symposium on Principles of Distributed Computing}{July 29-August 2, 2019}{Toronto, ON, Canada}
\acmBooktitle{2019 ACM Symposium on Principles of Distributed Computing (PODC '19), July 29-August 2, 2019, Toronto, ON, Canada}
\acmPrice{15.00}
\acmDOI{10.1145/3293611.3331590}
\acmISBN{978-1-4503-6217-7/19/07}

\begin{document}

\iflong
\title[Reconfigurable Atomic Transaction Commit (Extended Version)]
    {Reconfigurable Atomic Transaction Commit (Extended Version)}
\else

\title[Reconfigurable Atomic Transaction Commit]
    {Reconfigurable Atomic Transaction Commit}
\fi

\author{Manuel Bravo}
\affiliation{
  \institution{IMDEA Software Institute}
}

\author{Alexey Gotsman}
\affiliation{
  \institution{IMDEA Software Institute}            
}



\begin{abstract}
  Modern data stores achieve scalability by partitioning data into shards and
  fault-tolerance by replicating each shard across several servers. A key
  component of such systems is a Transaction Certification Service (TCS), which
  atomically commits a transaction spanning multiple shards. Existing TCS
  protocols require $2f+1$ crash-stop replicas per shard to tolerate $f$
  failures. In this paper we present atomic commit protocols that require only
  $f+1$ replicas and reconfigure the system upon failures using an external
  reconfiguration service. We furthermore rigorously prove that these protocols
  correctly implement a recently proposed TCS specification. We present
  protocols in two different models---the standard asynchronous message-passing
  model and a model with Remote Direct Memory Access (RDMA), which allows a
  machine to access the memory of another machine over the network without
  involving the latter's CPU. Our protocols are inspired by a recent FARM system
  for RDMA-based transaction processing. Our work codifies the core ideas of
  FARM as distributed TCS protocols, rigorously proves them correct and
  highlights the trade-offs required by the use of RDMA.
\end{abstract}

\let\oldaddcontentsline\addcontentsline
\def\addcontentsline#1#2#3{}
\maketitle

\section{Introduction}

Modern data stores are often required to manage massive amounts of data while
providing stringent transactional guarantees to their users. They achieve
scalability by partitioning data into independently managed {\em shards} (aka
{\em partitions}) and fault-tolerance by replicating each shard across a set of
servers~\cite{spanner,scatter,gdur}. Such data stores often use optimistic
concurrency control~\cite{wv}, where a transaction is first executed
speculatively, and the results (e.g., read and write sets) are then {\em
  certified} to determine whether the transaction can commit or must abort
because of a conflict with concurrent transactions. The certification is
implemented by a {\em Transaction Certification Service (TCS)}, which accepts a
stream of transactions and outputs decisions based on a given {\em certification
  function}, defining the concurrency-control check for the desired isolation
level. TCS is the most challenging part of transaction processing in systems
with the above architecture, since it requires solving a distributed agreement
problem among the replicated shards participating in the transaction. This
agreement problem has been recently formalized as the {\em multi-shot commit
  problem}~\cite{discpaper}, generalizing the classical {\em atomic commit
  problem}~\cite{dwork-skeen} to more faithfully reflect the requirements of
modern transaction processing systems (we review the new problem statement in
\S\ref{sec:tcs}).

Most existing solutions to the TCS problem require replicating each shard among
$2f+1$ replicas to tolerate $f$ crash-stop failures within each
shard~\cite{spanner,scatter,uw-inconsistent,mdcc}, which allows using a
replication protocol such as Paxos~\cite{paxos}.
This is expensive: if transaction data are written to all replicas of the
shard, only $f+1$ replicas are needed for the data to survive failures. Since,
in this case even a single replica failure will block transaction processing, to
recover we need to {\em reconfigure} the system, i.e., change its membership to
replace failed replicas with fresh ones. Unfortunately, processes concurrently
deciding to reconfigure the system need to be able to agree on the next
configuration; this reduces to solving consensus, which again requires $2f+1$
replicas~\cite{lower-bound}. The way out of this conundrum is to use a separate
{\em configuration service} with $2f+1$ replicas to perform consensus on the
configuration. In this way, we use $2f+1$ replicas only to store the small amount
of information about the configuration and $f+1$ replicas to store the actual
data. This {\em vertical approach}~\cite{vertical-paxos}, which layers
replication on top of a configuration service, has been used by a number of
practical systems~\cite{corfu,bigtable,farm}. It is particularly suitable for
deployment in local-area networks, where the configuration service can be
reached quickly.

In this paper we propose the first rigorously proven protocols for implementing
a TCS in a vertical system, with $f+1$ replicas per shard and an external
configuration service. We present protocols in two different models---the
standard asynchronous message-passing model (\S\ref{sec:cft}) and a model with
Remote Direct Memory Access (RDMA), which allows a machine to access the memory
of another machine over the network without involving the latter's CPU
(\S\ref{sec:rdmadiscussion}). Our protocols are parametric in the isolation
level provided, and we prove that they correctly implement the TCS specification
from the multi-shot commit problem~\cite{discpaper} (\S\ref{sec:correctness}).

Our work complements and takes its inspiration from a recent FARM
system~\cite{farm,farm2}---a transaction processing system that achieves
impressive scalability and availability by exploiting RDMA and the vertical
approach. FARM currently forms the core of a graph database used to serve some
of search queries in Microsoft Bing. It is a complex system that includes a
number of optimizations, both specific to RDMA and not. FARM's design was
presented without a rigorous proof of correctness, and it did not highlight
which features are motivated by the use of RDMA and which are inherent to the
vertical approach. Our work provides a theoretical complement to FARM: we codify
its core ideas as distributed transaction commit protocols and rigorously prove
them correct with respect to the TCS specification. By basing our protocols on a
principled footing, we are also able to provide better fault-tolerance
guarantees than FARM. Finally, by presenting two related protocols using message
passing and RDMA, we determine the trade-offs required by the use of RDMA.

In more detail, a straightforward way to implement TCS is using the classical
{\em two-phase commit (2PC)} protocol~\cite{2pc}. Since 2PC is not
fault-tolerant, we can make each shard simulate a reliable process in 2PC using
a replication protocol such as Paxos~\cite{spanner,scatter}. This vanilla
approach requires every 2PC action to be replicated using Paxos, which results
in a high latency (7 message delays to learn a decision on a
transaction~\cite{replicated-commit}) and a high load on Paxos leaders. To
improve on this, our protocol combines 2PC and Vertical
Paxos~\cite{vertical-paxos} into one coherent protocol, thereby minimizing the
latency and load on Paxos leaders. Upon a failure inside a shard, we use the
reconfiguration service to replace the failed replicas, as in Vertical
Paxos. This reconfiguration interacts nontrivially with the 2PC part of the
protocol: e.g., reconfiguration may lead to losing undecided transactions that
affected 2PC computations of decisions on other transactions---a behavior that
we nevertheless show to be correct. Finally, we show that the price of
exploiting RDMA to efficiently write transaction data to replicas is that
reconfiguration has to be performed globally, instead of per-shard: when
reconfiguring a shard, we have to ensure that the whole system is
aware of the configuration before activating it.




\section{Transaction Certification Service}
\label{sec:tcs}

\subparagraph*{Service interface and certification functions.} A {\em Transaction
  Certification Service (TCS)} is meant to be used in the context of
transactional processing systems with optimistic concurrency control~\cite{wv},
where transactions are first executed speculatively, and the results are
submitted for certification to the TCS. We start by reviewing its specification
proposed in~\cite{discpaper}.
Clients invoke the TCS using requests of the form $\Certify(t, \pl)$, where
$t \in \Txn$ is a unique transaction identifier and $\pl\in \Payload$ is the
transaction payload, which carries the results of the optimistic execution of
the transaction (e.g., read and write sets). Responses of the service are of the
form $\Decide(t, d)$, where $d \in \D=\{\ABORT, \COMMIT\}$. A TCS is specified
using a {\em certification function} $\cf: 2^{\Payload} \times \Payload \to \D$,
which encapsulates the concurrency-control policy for the desired isolation
level. The result $\cf(\setpayload, \pl)$ is the decision for the transaction
with payload $\pl$ given the set of payloads $\setpayload$ of the previously
committed transactions. We require $\cf$ to be {\em distributive} in the
following sense:
\begin{equation}
\label{distributive}
\forall \setpayload_1, \setpayload_2, \pl.\, \cf(\setpayload_1 \cup \setpayload_2, \pl) = \cf(\setpayload_1, \pl) \sqcap \cf(\setpayload_2, \pl),
\end{equation}
where $\sqcap$ is such that $\COMMIT \sqcap \COMMIT = \COMMIT$ and
$d \sqcap \ABORT = \ABORT$ for any $d$. This requirement is justified by the
fact that common definitions of $\cf(\setpayload, \pl)$ check $\pl$ for
conflicts against each transaction in $\setpayload$ separately.

As an example, consider a transactional system managing \emph{objects} from
$\Obj$ with values from $\Val$, where transactions can execute reads and writes
on the objects. The objects are associated with a totally ordered set $\Ver$ of
{\em versions}.
Then the payload of a transaction $t$ is a triple $\langle R, W, V_c
\rangle$. Here the \emph{read set} $R \subseteq \Obj \times \Ver$ is the set of
objects with their versions that $t$ read, which contains one version per
object. The \emph{write set} $W \subseteq \Obj \times \Val$ is the set of
objects with their values that $t$ wrote, which contains one value per
object. We require that any object written has also been read:
$\forall (x, \_) \in W.\, (x, \_) \in R$. Finally, the \emph{commit version}
$V_c \in \Ver$ is the version to be assigned to the writes of $t$. We require
this version to be higher than any of the versions read:
$\forall (\_, v) \in R.\, V_c > v$. Given this domain of transactions, the
following certification function encapsulates the classical concurrency-control
policy for serializability~\cite{wv}:
$\cf(\setpayload, \langle R, W, V_c\rangle)=\COMMIT$ iff none of the versions in
$R$ have been overwritten by a transaction in $\setpayload$, i.e.,
\begin{multline}
\label{serializability-cf}
\forall x, v.\, (x, v) \in R\implies \\
(\forall (\_, W', V'_c) \in \setpayload  .\, (x, \_) \in W' \implies V_c' \le v).
\end{multline}

\subparagraph*{TCS specification.}
We represent TCS executions using {\em histories}---sequences of \Certify\/ and
\Decide\/ actions such that every transaction appears at most once 
in \Certify, and each \Decide\/ is a response to exactly one preceding
\Certify. For a history $h$ we let $\act(h)$ be the set of actions in
$\hist$. For actions $a, a'\in \act(\hist)$, we write $a \prec_\hist a'$ when
$a$ occurs before $a'$ in $h$. A history $\hist$ is {\em complete}\/ if every
$\Certify$ action in it has a matching $\Decide$ action. A complete history is
{\em sequential}\/ if it consists of pairs of $\Certify$ and matching $\Decide$
actions. A transaction $t$ {\em commits} in a history $h$ if $h$ contains
$\Decide(t, \COMMIT)$. We denote by $\committed(\hist)$ the projection of $h$ to
actions corresponding to the transactions that are committed in $\hist$. For a
complete history $\hist$, a {\em linearization}\/ $\ell$ of
$\hist$~\cite{linearizability} is a sequential history such that $\hist$ and
$\ell$ contain the same actions and
\begin{multline*}
\forall t, t'.\, \Decide(t, \_) \prec_{\hist} \Certify(t', \_)
\implies \\
\Decide(t,\_) \prec_{\ell} \Certify(t', \_).
\end{multline*}
A complete sequential history $h$ is {\em legal} with respect to a certification
function $\cf$, if its decisions are computed according to $\cf$:
\begin{multline*}
\forall t,\pl, d.\, \Certify(t,\pl), \Decide(t, d) \in \act(h) \implies
\\
d = \cf(\{\pl' \mid \Certify(t',\pl')\in \act(h) \wedge {}\\
\Decide(t', \COMMIT) \prec_h \Decide(t, d)\}, \pl).
\end{multline*}
A history $\hist$ is {\em correct} with respect to $\cf$ if
$\hist \mid \committed(\hist)$ has a legal linearization. A TCS implementation
is {\em correct} with respect to $\cf$ if so are all its histories.

A TCS implementation satisfying the above specification can be readily used in a
transaction processing system. For example, consider the domain of transactions
defined earlier. A typical system based on optimistic concurrency control will
ensure that transactions submitted for certification only read versions written
by previously committed transactions. A history produced by such a system that
is correct with respect to certification function~(\ref{serializability-cf}) is
also serializable~\cite{discpaper}. Hence, a TCS correct with respect to this
certification function can indeed be used to implement serializability.

\subparagraph*{Shard-local certification functions.}  We are interested in TCS
implementations in systems where the data are partitioned into shards from a set
$\Shards$. In such systems TCS is usually implemented using a variant of the
classical {\em two-phase commit protocol (2PC)}~\cite{2pc}. In this protocol
each shard $s$ receiving a transaction for certification first {\em prepares}
it, i.e., performs a local concurrency-control check and accordingly votes to
commit or abort the transaction. The votes on the transaction by different
shards are aggregated, and the final decision is then distributed to all shards:
the transaction can commit only if all votes are commit. When a shard $s$ votes
on a transaction, it does not have information about all transactions in the
system, but only those that concern it. Hence, the votes are computed using not
the global certification function $f$, but \emph{shard-local certification
  functions}~\cite{discpaper}, which check for conflicts only on objects managed
by the shard and correspondingly take as parameters only the parts of the
transaction payloads relevant to the shard: for a payload $\pl$ we denote this
by $\pl\mid s$. For example, let $\Obj_s$ be the set of objects managed by a
shard $s$. For a payload $\pl = \langle R, W, V_c \rangle$ of the form given
above, we let $\pl \mid s = \langle R^s, W^s, V_c\rangle$, where
$R^s=\{(x,\_)\in R \mid x\in \Obj_s\}$ and
$W^s=\{(x,\_)\in W \mid x\in \Obj_s\}$. There are two shard-local functions,
$\cf_s : 2^{\Payload} \times \Payload \to \D$ and
$\cg_s : 2^{\Payload} \times \Payload \to \D$. As its first argument $\cf_s$
takes the set of shard-relevant payloads of transactions that previously
committed at the shard, and $\cg_s$ the set of such payloads for transactions
that have been prepared to commit. As their second argument, the functions take
the part of the payload of the transaction being certified relevant to the
shard. We require that these functions are distributive, similarly to
(\ref{distributive}).

For example, the shard-local certification functions for serializability are
defined as follows: $\cf_s(\setpayload, \langle R, W, V_c \rangle)=\COMMIT$ iff
\begin{equation}\label{cf-local-ser-fs}
\begin{array}{@{}l@{}l@{}l@{}l@{}}
\forall x \in \Obj_s.\, \forall v.\, {} &
(x, v) \in R \implies & \\[2pt]
&
(\forall \langle \_, W', V'_c\rangle \in \setpayload .\,
               (x, \_) \in W' \implies & V'_c \le v), \nonumber
\end{array}
\end{equation}
and $\cg_s(\setpayload, \langle R, W, V_c \rangle)=\COMMIT$ iff
\begin{equation}\label{cf-local-ser-gs}
\begin{array}{@{}l@{}l@{}l@{}}
\forall x \in \Obj_s.\, \forall v.\, {} &
((x, \_) \in R & \implies (\forall \langle \_, W', \_ \rangle \in \setpayload .\, (x, \_) \not\in W'))
\wedge {}
\\[2pt]
& ((x, \_) \in W & \implies (\forall \langle R', \_, \_ \rangle \in \setpayload .\, (x, \_) \not\in R')). \nonumber
\end{array}
\end{equation}
The function $\cf_s$ certifies a transaction $t$ against previously committed
transactions similarly to the certification function (\ref{serializability-cf}) for serializability, but taking
into account only the objects managed by the shard $s$. The function $\cg_s$
certifies $t$ against transactions prepared to commit, and its check is stricter
than that of $\cf_s$. In our example, the function $\cg_s$ aborts a transaction
$t$ if: {\em (i)} it read an object written by a transaction $t'$ prepared to commit;
or {\em (ii)} it writes to an object read by a transaction $t'$ prepared to
commit. This reflects the behaviour of typical implementations, which upon
preparing a transaction acquire read locks on its read set and write locks on
its write set, and abort the transaction if the locks cannot be acquired.

For a sharded TCS implementation to be correct, shard-local functions have to
{\em match} the global certification function, i.e., perform similar conflict
checks. 
We formalize the required conditions as follows. Assume a function
$\shards: \Txn \to 2^\Shard$ that determines the shards that need to certify a
transaction with a given identifier, which are usually the shards storing the
data the transaction accesses. We also assume a distinguished {\em empty}
payload $\varepsilon\in\Payload$ such that
$\forall s, \setpayload.\, \cf_s(L, \varepsilon) = \COMMIT$. For example, for a
payload $\pl =\langle R,W,\_\rangle$ of the form given above, $\pl=\varepsilon$
is such that $R=\emptyset$ and $W=\emptyset$. We require that for a transaction
$t\in\Txn$ with payload $\pl\in\Payload$, for each shard $s\not\in\shards(t)$,
we have $\pl\mid s=\varepsilon$. We further lift the $\mid$ operator to sets of
payloads: for any $\setpayload\subseteq\Payload$ we let
$(\setpayload\mid s) = \{(\pl\mid s)\mid \pl\in\setpayload\}$. Then we require
that global and local certification functions match as follows:
\begin{multline}
\label{local-global}
\forall \pl\in\Payload.\,\forall \setpayload\subseteq\Payload.\,
\cf(\setpayload, \pl)=\COMMIT \iff {} \\
\forall s\in \Shards.\, \cf_s((\setpayload \mid s),
(\pl \mid s))=\COMMIT.
\end{multline}

Finally, for each shard $s\in\Shards$, the two functions $\cf_s$ and $\cg_s$ are
required to be related to each other as follows~\cite{discpaper}:
\begin{gather}
\label{g-implies-f}
\forall \pl\in\Payload.\, \setpayload\subseteq\Payload.\, \cg_s(L, \pl) = \COMMIT
{\implies} \cf_s(L, \pl) = \COMMIT;
\\[2pt]
\forall \pl, \pl'\in\Payload.\, \cg_s(\{\pl\}, \pl') \,{=}\,
\COMMIT {\implies} \cf_s(\{\pl'\}, \pl) \,{=}\, \COMMIT.
\label{g-conflict-f}
\end{gather}
Property~(\ref{g-implies-f}) requires the conflict check performed by $g_s$ to
be no weaker than the one performed by $f_s$. Property~(\ref{g-conflict-f})
requires a form of commutativity: if a transaction with payload $l'$ is allowed
to commit after a still-pending transaction with payload $l$, then the latter
would be allowed to commit after the former.


\section{Atomic Commit Protocol}
\label{sec:cft}

\subparagraph*{System model.}  We consider an asynchronous message-passing
system consisting of a set of processes $\Proc$ which may fail by crashing,
i.e., permanently stopping execution. We assume that processes are connected by
reliable FIFO channels: messages are delivered in FIFO order, and messages
between non-faulty processes are guaranteed to be eventually delivered. A
function $\client: \Txn \to \Proc$ determines the client process that issued a
given transaction. 

Each shard $s\in\Shards$ is managed by a group of {\em replica} processes, whose
membership can change over time. For simplicity, we assume that the groups of
replica processes managing different shards are disjoint.
Each shard moves through a
sequence of \emph{configurations}, determining its
membership. \emph{Reconfiguration} is the process of changing the configuration
of a shard. In our protocols reconfiguration is initiated by a replica when it
suspects another replica of failing: for simplicity we do not expose it in the
TCS interface.
Every member of a shard in a given configuration is either the {\em leader} of
the shard or a {\em follower}. A configuration of a shard $s$ is then a tuple
$\langle \vballot, \vm, p_l\rangle$ where $\vballot$ is the {\em epoch}
identifying the configuration, $\vm\in 2^\Proc$ is the set of processes that
manage $s$ at $\vballot$, and $p_l\in\vm$ is the leader of $s$ at $\vballot$.

Configurations are stored in an external \emph{\external{} (CS)}, which for
simplicity we assume to be a reliable process. In practice, this service may be
implemented using Paxos-like replication over $2f+1$ processes out of which at
most $f$ can fail (as done in systems such as Zookeeper~\cite{zookeeper}). The
configuration service stores the configurations of all shards and provides three
operations. An operation {\tt
  compare\_and\_swap}$(s, e, \langle e', \vm, p_l\rangle)$ succeeds if the epoch
of the last stored configuration of $s$ is $e$; in this case it stores the
provided configuration with a higher epoch $e' > e$.
Operations {\tt get\_last}$(s)$ and {\tt get}$(s, e)$ respectively return the
last configuration of $s$ and the configuration of $s$ associated with a given
epoch $e$.

\begin{figure*}[t]
\center
\includegraphics[width=.98\textwidth]{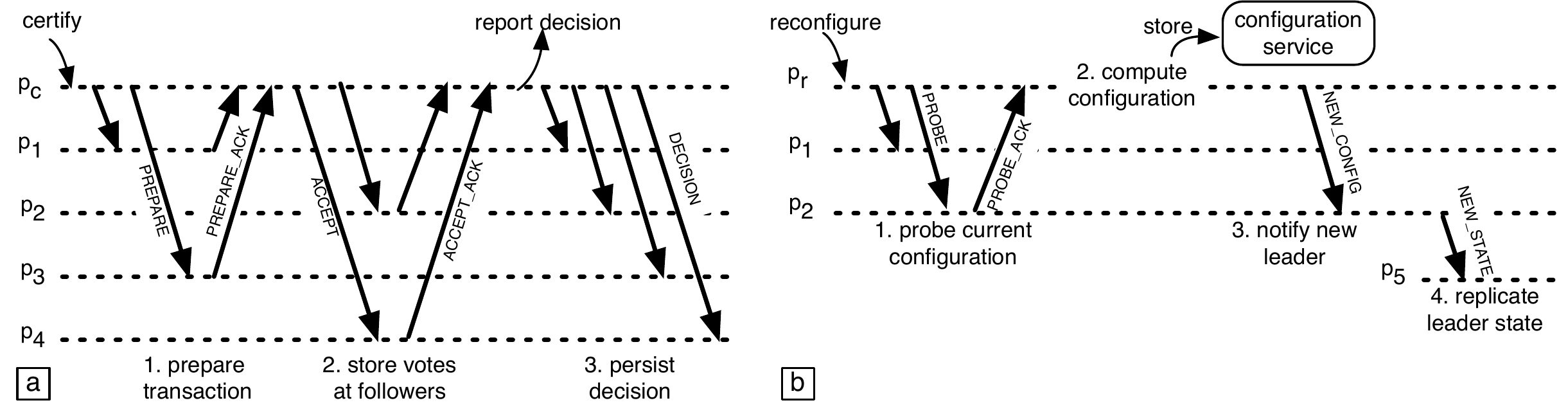}
\caption{\rm Illustrations of the behavior of the atomic commit protocol during
  (a) failure-free operation; (b) reconfiguration. In both cases $p_1$ is the
  initial leader of shard $s_1$ and $p_2$ its follower; $p_3$ is the leader of
  shard $s_2$ and $p_4$ its follower. In (b), after the reconfiguration of
  $s_1$, process $p_2$ becomes the leader of $s_1$ and a fresh process $p_5$
  becomes its follower.}
\label{fig:flow}
\end{figure*}

\subparagraph*{Protocol preliminaries.}  We give the pseudocode of our protocol
in Figure~\ref{fig:protocol}, illustrate its message flow in
Figure~\ref{fig:flow} and summarize the key invariants used in its proof of
correctness in Figure~\ref{fig:inv}. The protocol weaves together the two-phase
commit protocol across shards~\cite{2pc} and a Vertical Paxos-based
reconfiguration protocol within each shard~\cite{vertical-paxos}. At any given
time, a process participates in a single configuration of the shard it belongs
to. The process stores the information about this configuration as well as those
of other shards in several arrays: configuration epochs are stored in an array
$\ballot \in \Shards \to \mathbb{N}$, the current members in
$\M \in \Shards \to 2^{\Proc}$, and the current leader in
$\leader \in \Shards \to \Proc$. The entries for the shard the process belongs
to give the configuration the process is in; the other entries maintain
information about the configurations of the other shards. A $\status$ variable
at a process records whether it is a $\LEADER$, a $\FOLLOWER$ or is in a special
$\RECOVERING$ state used during reconfiguration. Each process keeps track of the
status of transactions in an array $\phase$, whose entries initially store
$\START$. The transaction status changes to $\ACCEPTED$ when the shard
determines its vote and to $\DECIDED$ when a final decision on the transaction
is reached.



\begin{figure*}
\begin{tabular}{@{}l@{ }|@{\quad\quad}l@{}}
\begin{minipage}[t]{9.5cm}
\removelatexerror
\begin{algorithm}[H]
  \SubAlgo{\Fun \Certify($t, \pl$)}{\label{alg:certify}
    \ForAll{$s\in \shards(t)$}{
    \Send $\PREPARE(t, (\pl\mid s))$ \KwTo $\leader[s]$;\label{alg:client-send}
    }
  }

  \smallskip
  \smallskip
  \smallskip

  \SubAlgo{\WhenReceived $\PREPARE(t, \pl)$ {\bf from $p_j$}}{\label{alg:prepare} 
    \textbf{pre:} $\status = \LEADER$\; \label{alg:check-leader} 
    \uIf{$\exists k.\, t = \txn[k]$}{\label{line:unique_accept}
        \Send $\PREPAREACK(\ballot[s_0], s_0, k, \txn[k],$\\
        \nonl \hspace{2.58cm}$\payload[k], \vote[k])$ \KwTo $p_j$}
    \Else{\label{alg:certify-first}
      $\nextv \leftarrow \nextv + 1$\;\label{line:next}
      $(\txn,\phase)[\nextv] \leftarrow  (t,\ACCEPTED)$\;
    \uIf{$\pl\neq \bot$}{
      $\vote[\nextv] \leftarrow f_{s_0}(\setpayload_1, \pl) \sqcap
      g_{s_0}(\setpayload_2,\pl)$\; \label{alg:compute-vote}
      $\payload[\nextv] \leftarrow \pl$\;
    }
    \Else{\label{line:else-abort}
      $\vote[\nextv] \leftarrow \ABORT$\;\label{alg:forced-vote}
      $\payload[\nextv] \leftarrow \varepsilon$;
    }
    \Send $\PREPAREACK(\ballot[s_0], s_0, \nextv, t, $\\
    \nonl \hspace{2.58cm}$ \payload[\nextv], \vote[\nextv])$ \KwTo $p_j$;
    }
  }

 \smallskip
  \smallskip
  \smallskip

\SubAlgo{\WhenReceived $\PREPAREACK(\vballot, s, k, t, \pl,
    d)$}{\label{alg:receive-prepareack}
    \textbf{pre:} $\ballot[s] = \vballot$\;\label{alg:coord-check}
    \Send $\ACCEPT(\vballot, k, t, \pl, d)$ \KwTo $\M[s]\setminus \leader[s]$;\label{alg:send-accept}
    }

 \smallskip
  \smallskip
  \smallskip

\SubAlgo{\WhenReceived $\ACCEPT(\vballot, k, t, \pl, d)$ {\bf from $p_j$}}{ \label{alg:receive-accept}
    \textbf{pre:} $\status = \FOLLOWER \wedge 
    \ballot[s_0] = \vballot$\; \label{alg:check-accept} 
    \If{$\phase[k] =$ {\normalfont \START}}{%
       $(\txn,\payload,\vote,\phase)[\nextv] \leftarrow  (t,\pl, d, \ACCEPTED)$\; \label{alg:accept-vote}
    }\label{alg:accept-store} \label{alg:accept-newtx}
    \Send $\ACCEPTACK(s_0, \vballot, k, t, d)$ \KwTo $p_j$; \label{alg:accept-ack}
  }

  \smallskip
  \smallskip
  \smallskip

  \SubAlgo{{\bf when for every $s \in \shards(t)$ received an
      $\ACCEPTACK(s, \ballot[s], k_s, t, d_s)$ from every $p_j\in
      \M[s]\setminus \leader[s]$}}{\label{alg:receive-acceptack}
    \Send $\DECISION(t, \bigsqcap_{s \in \shards(t)} d_s)$ 
    {\bf to $\client(t)$}\; \label{alg:send-decision-client}
    \ForAll{$s \in \shards(t)$}{\label{alg:send-decision-groups1}%
      \Send $\DECISION(\ballot[s], k_s, \bigsqcap_{s \in \shards(t)} d_s)$ \\
      \nonl \quad \KwTo $\M[s]$;\label{alg:send-decision-groups2}}
  }

  \smallskip
  \smallskip
  \smallskip

  \SubAlgo{\WhenReceived $\DECISION(\vballot, k, d)$}{ \label{alg:receive-decision}
    \textbf{pre:} $\status \in \{\LEADER, \FOLLOWER\} \wedge \ballot[s_0] \geq
    \vballot$\; \label{alg:check-decision}
    $(\decision,\phase)[k] \leftarrow (d, \DECIDED)$\; \label{alg:decision-store}\label{alg:decision-phase}
}

\smallskip
  \smallskip
  \smallskip

\SubAlgo{\Fun ${\tt reconfigure}(s)$}{ \label{alg:recover}
    \textbf{pre:} $\probing=\FALSE$\;
    $\probing\gets \TRUE$\;
    $\langle\pballot, \pmembers, \_ \rangle \gets$ {\tt get\_last}($s$) {\bf at}
    $\zk$\; \label{alg:read-config} 
    $\rballot \gets \pballot+1$\;
    $\rshard \gets s$\;
    \Send $\PROBE(\rballot)$ \KwTo $\pmembers$\;\label{alg:sendprobe1}
    }

\end{algorithm}
\end{minipage}
&
\begin{minipage}[t]{8.5cm}
\removelatexerror
\begin{algorithm}[H]

\SubAlgo{\WhenReceived $\PROBE(\vballot)$ \KwFrom
    $p_j$}{\label{alg:probe}
    \textbf{pre:} $\vballot \geq\cballot$\;
    $\status = \RECOVERING$\;\label{line:statusreconf}
    $\cballot\gets \vballot$\;\label{line:probe-set-newepoch}
    \Send $\PROBEACK(\recovery, \vballot, s_0)$ \KwTo $p_j$\;
    }

\smallskip
  \smallskip
  \smallskip
  
\SubAlgo{\WhenReceived $\PROBEACK(\TRUE, \rballot,$\\
  \nonl\hspace{3.6cm} $\rshard)$ \KwFrom $p_j$}{\label{alg:end-probing} 
      \textbf{pre:} $\probing=\TRUE$\;
     $\probing\gets \FALSE$\;
     $\vm\gets$ {\tt compute\_membership}()\;\label{line:members}
    {\bf var} $r\gets$ {\tt compare\_and\_swap}($\rshard,$ \label{line:writezk}\\
    \nonl \quad $\rballot-1, \langle \rballot, \vm, p_j \rangle$) {\bf at} $\zk$\; 
    \lIf{$r$}{\Send $\NEWCONFIG(\rballot, \vm)$ \KwTo $p_j$}
}
  \smallskip
  \smallskip
  \smallskip

\SubAlgo{{\bf non-deterministically when received} $\PROBEACK(\FALSE,
  \rballot, \rshard)$ {\bf from}
  $p_j \in \pmembers$ {\bf and no} $\PROBEACK(\TRUE, \rballot, \rshard)$}{\label{alg:advance-probing}
  \textbf{pre:} $\probing=\TRUE$\;
  $\pballot\gets\pballot-1$\;
  $\pmembers \gets$ {\tt get}($\rshard,$\\
  \nonl\hspace{3.3cm}$\pballot$) {\bf at} $\zk$\;
   \Send $\PROBE(\rballot)$ \KwTo $\pmembers$\;\label{alg:sendprobe2}
}

\smallskip
  \smallskip
  \smallskip



\SubAlgo{\WhenReceived $\NEWCONFIG(\cballot, M)$ \KwFrom
    $p_j$}{\label{alg:receivenewconfig-leader}
    $\status = \LEADER$\;\label{alg:set-status-leader}
    $(\ballot,\M,\leader)[s_0]\gets (\vballot,M,p_i)$\;
    $\nextv\gets \max\{k \mid \phase[k] \not= \START\}$\;\label{line:setnext}
    \Send $\REPLICATELOGS(\cballot, M, \txn, \payload, \vote, $\\
    \nonl \hspace{2.25cm}$\dec, \phase)$ \KwTo
    $M\setminus p_i$\;\label{alg:send-newview} 
}

  \smallskip
  \smallskip
  \smallskip

\SubAlgo{\WhenReceived $\REPLICATELOGS(\vballot, M, \vtxn, \vpayload, \vvote,$\\ 
    \nonl \hspace{3.55cm} $ 
    \vdecision, \vphase)$ \KwFrom
    $p_j$}{\label{alg:newview}
    \textbf{pre:}
    $\vballot\geq\cballot$\; \label{alg:check-replicatelogs}
    $\recovery\gets \TRUE$\;
    $\status = \FOLLOWER$\;
    $(\ballot, \M,\leader)[s_0]\gets (\vballot,M,p_j)$\;
    $(\txn, \payload, \vote,\decision,\phase) \leftarrow (\vtxn, \vpayload, \vvote, \vdecision, \vphase)$\; \label{alg:new-vote}
    }

  \smallskip
    \smallskip
  \smallskip

\SubAlgo{\WhenReceived $\CONFIGCHANGE(s, \vballot, M, p_l)$
  {\bf from} $\zk$}{\label{alg:configchange}
  \textbf{pre:} $\ballot[s]<\vballot\wedge s\neq s_0$\;
    $(\ballot, \M,\leader)[s]\gets (\vballot, M,p_l)$\;
}

 \smallskip
  \smallskip
  \smallskip

\SubAlgo{\Fun ${\tt retry}(k)$}{ \label{alg:retry}
    \textbf{pre:} $\phase[k] = \ACCEPTED$\;
    \ForAll{$s\in \shards(\txn[k])$}{
    \Send $\PREPARE(\txn[k], \bot)$ \KwTo $\leader[s]$;\label{alg:choose-coord}
  }}

\end{algorithm}
\end{minipage}
\end{tabular}
\\
\caption[center]{\rm Atomic commit protocol at a process $p_i$ in a shard $s_0$. At line~\ref{alg:compute-vote} we let\\
  $\setpayload_1=\{\payload[k] \mid k < \nextv \wedge\phase[k] =$
  {\normalfont \DECIDED}$ {} \wedge
\decision[k] = $ {\normalfont \COMMIT}$\}$;\\
$\setpayload_2=\{\payload[k] \mid k < \nextv \wedge \phase[k] =$ {\normalfont \ACCEPTED}${} \wedge
\vote[k] = $ {\normalfont \COMMIT}$\}$.}
\label{fig:protocol}
\end{figure*}

\begin{figure}
\begin{enumerate}[leftmargin=10pt, label={\arabic*.}, ref={\arabic*}]
\item\label{inv:synchrony} If a process $p_i$ receives
  $\ACCEPT(\vballot, k, t, \pl, d)$ and replies with
  $\ACCEPTACK$, then after this and while $\ballot[s]=\vballot$ at $p_i$,
  we have $\prefix{\txn}{k} \prec \prefix{\vtxn}{k}$, 
  $\prefix{\vote}{k} \prec \prefix{\vvote}{k}$ and
  $\prefix{\payload}{k} \prec \prefix{\vpayload}{k}$, where $\vtxn$,
  $\vvote$ and $\vpayload$ were
  the values of the arrays $\txn$, $\vote$ and $\payload$ at the leader of $s$ at $\vballot$
  when it sent the corresponding message $\PREPAREACK(\vballot, s, k,
  t, \pl, d)$.


\item\label{inv:main} Assume that all followers in $s$ at $\vballot$ received
  $\ACCEPT(\vballot, k, t, \pl, d)$ and responded to it with $\ACCEPTACK$, and
  at the time the leader of $s$ at $\vballot$ sent the corresponding message
  $\PREPAREACK(\vballot, s, k, t, \pl, d)$ it had $\prefix{\txn}{k} = \vtxn$,
  $\prefix{\vote}{k} = \vvote$  and
  $\prefix{\payload}{k} = \vpayload$. Whenever at a process in $s$ we have
  $\ballot[s] = \vballot' > \vballot$, we also have $\prefix{\txn}{k} \prec
  \vtxn$, $\prefix{\vote}{k} \prec \vvote$ and
  $\prefix{\payload}{k} \prec \vpayload$.

\item\label{inv:probe} After a process receives $\PROBE(\vballot)$ and
  replies with $\PROBEACK(\_, \vballot, s)$, it will never send
  $\ACCEPTACK(s, \vballot', \_, \_, \_)$ with $\vballot'<\vballot$.

\item\label{inv:consensus} 
  \begin{enumerate}[leftmargin=0.5cm]
  \item For any messages $\DECISION(\_, k, d_1)$ and $\DECISION(\_, k, d_2)$
    sent to processes in the same shard, we have $d_1 = d_2$.
  \item For any messages $\DECISION(t, d_1)$ and $\DECISION(t, d_2)$ sent, we
    have $d_1 = d_2$.
  \end{enumerate}

\item\label{inv:membership} Assume that all followers in $s$ at $\vballot$
  received $\ACCEPT(\vballot, k, t, \pl, d)$ and replied with $\ACCEPTACK$. Let
  the process $p_i$ be a member of $s$ at $\vballot'$ such that
  $\vballot'<\vballot$. If $p_i$ is not a member of $s$ at $\vballot$ then $p_i$
  cannot be member of $s$ at any $\vballot''>\vballot$.

\end{enumerate}
\caption{\rm Key invariants of the atomic commit protocol. Given a sequence
  $\alpha$, we let $\length({\alpha}) = \max\{k \mid \alpha[k] \not= \bot\}$ and
  let $\prefix{\alpha}{k}$ be the prefix of $\alpha$ of length $k$. Given a
  second sequence $\beta$, we let $\beta\prec \alpha$ if
  \mbox{$\length(\alpha) \,{=}\, \length(\beta) \,{\wedge}\, (\forall k \,{\leq}\, \length(\beta).\,
  \beta[k] \,{\neq}\, \bot {\implies} \beta[k] \,{=}\, \alpha[k])$.}}
\label{fig:inv}
\end{figure}

\subparagraph*{Failure-free case.}  A client submits a transaction for
certification by calling the {\tt certify} function at any replica process,
which will serve as the {\em coordinator} of the transaction
(line~\ref{alg:certify}). The function takes as arguments the transaction's
identifier and its payload. The transaction coordinator first sends a
$\PREPARE$ message to the leaders of the relevant shards, which includes the
payload part for each shard (line~\ref{alg:client-send}). The leader
of a shard arranges all transactions it receives into a total
\emph{certification order}, which the leader stores in an array
$\txn \in \mathbb{N} \to \Txn$; a $\nextv \in \mathbb{Z}$ variable points to the
last filled slot in the array. When the leader receives a $\PREPARE$ message for
a transaction for the first time (line~\ref{alg:certify-first}), it appends the
transaction to the certification order, stores the transaction's payload in an
array $\payload \in \mathbb{N} \to \Payload$, and sets the transaction's phase
to $\ACCEPTED$. It then computes a vote on the transaction and stores it in an
array $\vote \in \mathbb{N} \to \{\COMMIT, \ABORT\}$
(line~\ref{alg:compute-vote}). The vote is computed using the shard-local
certification functions $\cf_{s_0}$ and $\cg_{s_0}$ to check for conflicts
against transactions that have been previously committed or prepared to commit;
the results are combined using the $\sqcap$ operator, so that the transaction
can commit only if both functions say so. We defer the description of the cases
when the leader has previously received the transaction in the $\PREPARE$
message (line~\ref{line:unique_accept}) and when the payload in the message is
an undefined value $\bot$ (line~\ref{line:else-abort}).

Our protocol next replicates the leader's decision and the transaction payload
at the followers. Instead of having the leader to do this directly, the protocol
delegates this task to the coordinator of the transaction. This design is used
by practical systems, such as Corfu~\cite{corfu} and FARM~\cite{farm}, since it
minimizes the load on the leaders, which are the main potential performance
bottleneck. Instead, the network-intensive task of persisting transactions at
multiple followers is spread among a number of different transaction
coordinators. As we explain in the following, this optimization interacts in
a nontrivial way with transaction certification. In more detail, after preparing
a transaction the leader sends a $\PREPAREACK$ message to the coordinator of the
transaction, which carries the leader's epoch, the transaction identifier, its
position in the certification order, the payload, and the vote
(line~\ref{alg:send-accept}).
Upon receiving the $\PREPAREACK$ message (line~\ref{alg:receive-prepareack}),
the coordinator 
forwards the data from the $\PREPAREACK$ message to the followers in an $\ACCEPT$ message.

A process handles an $\ACCEPT$ message only if it participates in the
corresponding epoch (line~\ref{alg:check-accept}). The process stores the
transaction identifier, its payload and vote, and advances the transaction's
phase to $\ACCEPTED$. It then sends an $\ACCEPTACK$ message to the coordinator
of the transaction, confirming that the process has accepted the transaction and
the vote. The certification order at a follower is always a prefix with zero or
more holes of the certification order at the leader of the epoch the follower is
in, as formalized by Invariant~\ref{inv:synchrony} (Figure~\ref{fig:inv}). The
holes in the prefix arise from the lack of FIFO ordering in the communication
between the leader of a given epoch and its followers, as the $\ACCEPT$ message
for a given transaction is sent to the followers by the coordinator of the
transaction and not directly by the leader.

The coordinator of a transaction $t$ acts once it receives $\ACCEPTACK$ messages
for $t$ from every follower of its shards $s \in \shards(t)$
(line~\ref{alg:receive-acceptack}); it determines this using the configuration
information it stores for every shard. The coordinator computes the final
decision on $t$ using the $\sqcap$ operator on the votes of each
involved shard:
the transaction can commit if all
votes are commit. The coordinator then sends the final decision in $\DECISION$
messages to the client and to each of the relevant shards.
When a process receives a decision for a transaction
(line~\ref{alg:receive-decision}), it stores the decision and advances the
transaction's phase to $\DECIDED$. In a realistic implementation, at this point
the process would also upcall into the transaction processing system running at
its server, to inform it about the decision and allow it to apply the
transaction's writes to the database if the decision is to commit.



In the absence of failures, our protocol allows the client to learn a decision
on a transaction in 5 message delays, instead of 7 required by vanilla protocols
that use Paxos as a black-box~\cite{spanner,scatter}. We can further reduce this
to 4 by co-locating the client with the transaction coordinator. The protocol
also minimizes the load on Paxos leaders, which are the main potential
bottleneck: each involved leader only has to receive one $\PREPARE$ and one
$\DECISION$ message, and send one $\PREPAREACK$ message.

\subparagraph*{Reconfiguration.} When a failure is suspected in a shard $s$, any
process can initiate a reconfiguration of the shard to replace failed
replicas. Reconfiguration is done only in the affected shard, without disrupting
others. It aims to preserve Invariant~\ref{inv:main}, which is key in proving
the correctness of the protocol. This assumes that all followers in $s$ at an
epoch $\vballot$ have received $\ACCEPT(\vballot, k, t, \vpload, d)$ and
responded to it with $\ACCEPTACK$; in this case we say that the transaction $t$
has been {\em accepted} at shard $s$. The invariant guarantees that the accepted
transaction $t$ will persist in epochs higher than $\vballot$; this is used to
prove that the protocol computes a unique decision on each transaction. The
invariant also guarantees that the entries preceding $t$ in the certification
order in epochs higher than $\vballot$ may only contain the votes that the
leader of $s$ at epoch $\vballot$ took into account when computing the vote $d$
on $t$ (some of these votes may be missing due to the lack of FIFO order in the
communication between the leader and its followers). This property is necessary
to guarantee that the protocol computes decisions according to a single global
certification order, as required by the TCS specification. 

To ensure Invariant~\ref{inv:main}, a process performing reconfiguration first
{\em probes} previous configurations to determine which processes are still
alive and to find a process whose state contains all transactions previously
accepted at the shard, which will serve the new leader. The new leader then
transfers its state to the members of the new configuration, thereby {\em
  initializing} them. A variable $\recovery\in \{\TRUE, \FALSE\}$ at a process
records whether it has ever been initialized. Our protocol guarantees that a
shard can become {\em operational}, i.e., start accepting transactions, only
after all its members have been initialized.

The probing phase is complicated by the fact that there may be a series of
failed reconfiguration attempts, where the new leader fails before initializing
all its followers. Hence, probing requires traversing epochs from the current
one down, skipping epochs that are not operational. Probing selects as the new
leader the first initialized process it encounters during this traversal; we can
show that this process is guaranteed to know about all transactions accepted at
the shard, and thus making it the new leader will preserve
Invariant~\ref{inv:main} (\S\ref{sec:correctness}).

In more detail, a process $p_r$ initiates a reconfiguration of a shard $s$ by
calling ${\tt reconfigure}(s)$ (line~\ref{alg:recover}). The process picks an
epoch number $\rballot$ higher than the epoch of $s$ stored in the \external and
then starts the probing phase, as marked by the flag $\probing$. The process
$p_r$ keeps track of the shard being reconfigured in $\rshard$, the epoch being
probed in $\pballot$ and the membership of this epoch in $\pmembers$. The
process initializes these variables when it first reads the current
configuration from the \external{} (line~\ref{alg:read-config}). It then sends a
$\PROBE$ message to the members of the current configuration, asking them to
join the new epoch $\rballot$. Upon receiving a $\PROBE(e)$ message
(line~\ref{alg:probe}), a process first checks that the proposed epoch is equal
or higher than the highest epoch it has ever been asked to join, which the
process stores in $\cballot$ (we always have $\ballot[s] \le \cballot$ at a
process in $s$). In this case, the process sets $\cballot$ to $e$ and changes
its status to $\RECOVERING$, which causes it to stop transaction processing. It
then replies to $p_r$ with a $\PROBEACK$ message, which indicates whether it has
been previously initialized or not. If $p_r$ finds a process that has previously
been initialized, and hence can serve as the new leader, $p_r$ ends probing
(line~\ref{alg:end-probing}). If $p_r$ does not find such a process in the epoch
$\pballot$ and receives at least one reply $\PROBEACK$ from a process that has
not been initialized (line~\ref{alg:advance-probing}), $p_r$ can conclude that
the epoch $\pballot$ is not operational and will never become such, because it
has convinced at least one of its members to join the new epoch; this is
formalized by Invariant~\ref{inv:probe}.
In this case $p_r$ starts probing the preceding epoch. Since no transactions
could have been accepted at the epoch $\pballot$, picking a new leader from an
earlier epoch will not lose any accepted transactions and thus will not violate
Invariant~\ref{inv:main}.

Once the probing finds a new leader $p_j$ for the shard $s$
(line~\ref{alg:end-probing}), the process $p_r$ computes the membership of the
new configuration using a function {\tt compute\_membership}
(line~\ref{line:members}). We do not prescribe a particular implementation of
this function, except that the new membership must contain the new leader $p_j$
and may only contain the processes that replied to probing or fresh
processes. The latter can be added to reach the desired level of fault
tolerance. Once the new configuration is computed, $p_r$ attempts to store it in
the \external using a compare-and-swap operation. This succeeds only if the
current epoch is still the epoch from which $p_r$ started probing, which means
that no concurrent reconfiguration occurred while $p_r$ was probing. In this
case, $p_r$ sends a $\NEWCONFIG$ message with the new configuration to the new
leader of $s$.

When the new leader of $s$ receives the $\NEWCONFIG$ message
(line~\ref{alg:receivenewconfig-leader}),
it sets $\nextv$ to the length of its sequence of transactions, $\ballot[s]$ to
the new epoch and $\status$ to $\LEADER$, which allows it to start processing
new transactions. It then sends a $\REPLICATELOGS$ message to its followers,
containing its state. Upon receiving this message (line~\ref{alg:newview}), a
process overwrites its state with the one provided, changes its status to
$\FOLLOWER$, and sets $\recovery$ to $\TRUE$. As part of the state update, the
process also updates its epoch $\ballot[s_0]$ to the new one. Hence, the process
will not accept transactions from the new leader until it receives the
$\REPLICATELOGS$ message.


When a new configuration of a shard $s$ is persisted in the \external, the
service sends it in a $\CONFIGCHANGE$ message to the members of shards other
than $s$. A process updates the locally stored configuration upon receiving this
message (line~\ref{alg:configchange}).

\subparagraph*{Coordinator recovery.}  If a process that accepted a transaction
$t$ does not receive the final decision on it, this may be because the
coordinator of $t$ has failed. In this case the process may decide to become a
new coordinator by executing a {\tt retry} function (line~\ref{alg:retry}). For
this, the process just sends a $\PREPARE(t,\bot)$ message to the leaders of the
shards of $t$, carrying a special undefined value $\bot$ as the payload. If a
leader receiving $\PREPARE(t,\bot)$ has already certified $t$, it re-sends the
corresponding $\PREPAREACK$ message to the new coordinator, including the
transaction payload and vote (line~\ref{line:unique_accept}). Otherwise, if the
leader does not have the payload of $t$, it prepares the transaction as aborted
and with an empty payload $\varepsilon$ (line~\ref{alg:forced-vote}). In either
case, the new coordinator will finish processing the transaction as usual. The
above case when the transaction is aborted because the leader of a shard does
not know its payload may arise when the old coordinator crashed in between
sending $\PREPARE$ messages to different shards. Note that if the old
coordinator was suspected spuriously and will try later to submit the
transaction to a shard where it was aborted, it will just get a $\PREPAREACK$
message with an $\ABORT$ vote.


Our protocol allows any number of processes to become coordinators of a
transaction at the same time. Nevertheless, the protocol ensures that they will
all reach the same decision, even in case of reconfigurations. We formalize this
by Invariant~\ref{inv:consensus}: part (a) ensures an agreement on the decision
on the $k$-th transaction in the certification order at a given shard; part (b)
ensures a system-wide agreement on the decision on a given transaction $t$. The
latter part establishes that the protocol computes a unique decision on each
transaction. Invariant~\ref{inv:consensus} is proved as a corollary of
Invariant~\ref{inv:main}.




\subparagraph*{Losing undecided transactions.} Recall that our protocol uses the
optimization that delegates persisting transactions at followers to
coordinators~\cite{corfu,farm}. We now highlight how this optimization interacts
with transaction certification. Because of the optimization, transactions
prepared by a leader of a shard $s$ can be persisted at followers out of
order. For example, $t_2$ may follow $t_1$ in the certification order at the
leader, but may be persisted at followers first.
If now the leader of $s$ and the coordinator of $t_1$ crashes before $t_1$ is
persisted at followers, $t_1$ will be lost forever, something that is allowed by
Invariant~\ref{inv:main} (due to the use of $\prec$). In this case we lose a
transaction $t_1$ on the basis of which the vote on the transaction $t_2$ was
computed (e.g., the payload $l_1$ of $t_1$ was in $L_2$ when the vote on $t_2$
was computed at line~\ref{alg:compute-vote}). This does not violate correctness,
since the vote on $t_2$ makes sense also in the context excluding $t_1$: due to
distributivity of certification functions (\S\ref{sec:tcs}), if $t_2$ was
allowed to commit in the presence of $t_1$ ($f_s(\{l_1\}, l_2) = \COMMIT$), it
can also commit in its absence ($f_s(\emptyset, l_2) = \COMMIT$). Note that in
this case a decision on $t_1$ could not have been exposed to the client:
otherwise $t_1$ could not get lost due to Invariant~\ref{inv:main}. Also note
that, since we assume the transaction execution component produces payloads with
read-sets containing only values written by committed transactions
(\S\ref{sec:tcs}), in the above case $t_2$ could not have read a value written
by $t_1$.


\section{Correctness}
\label{sec:correctness}

The next theorem states the safety of our protocol, showing that it implements
the TCS specification.
\begin{theorem}
\label{thm:conc}
  A transaction certification service implemented using the protocol in
  Figure~\ref{fig:protocol} is correct with respect to a
  certification function $\cf$ matching the shard-local certification functions
  $\cf_s$ and $\cg_s$.
\end{theorem}
We defer the proof to~\tr{app:proof-conc} and only sketch the proof of the key
Invariant~\ref{inv:main}. This relies on auxiliary
Invariant~\ref{inv:membership}, which we prove first.

\subparagraph*{Proof sketch for Invariant~\ref{inv:membership}.} We prove the
invariant by induction on $\vballot''$. Assume that the invariant holds for all
$\vballot'' < \vballot^*$. We now show it for $\vballot'' = \vballot^*$. The
members of $s$ at $\vballot^*$ are computed at line~\ref{line:members} by a
reconfiguring process $p_r$ using the {\tt compute\_membership} function, which
returns either fresh processes or processes that responded to $p_r$'s probing.
Since $p_i$ was a member of $s$ at $e' < e^*$, it is not fresh; then by
assumptions on {\tt compute\_membership} $p_i$ must have received
$\PROBE(\vballot^*)$ from $p_r$ and replied with $\PROBEACK(\_, \vballot^*,
s)$. The process $p_r$ starts probing at epoch $\vballot^*-1$
and ends it upon receiving a $\PROBEACK(\TRUE, \vballot^*, s)$ message. By the
induction hypothesis, $p_i$ is not a member of $s$ at any epoch from
$\vballot^*-1$ down to $\vballot+1$. Hence, if the probing stops before reaching
$\vballot$, then $p_i$ will not be a member of $s$ at $e^*$, as required. Assume
now that the probing reaches $\vballot$. By Invariant~\ref{inv:probe}, each
follower in $s$ at $\vballot$ must have sent $\ACCEPTACK(s, \vballot, t)$ before
receiving $\PROBE(\vballot^*)$. Then any member of $s$ at $\vballot$ receiving
$\PROBE(\vballot^*)$ will have $\recovery=\TRUE$. Hence, if any member of $s$ at
$\vballot$ replies with $\PROBEACK(\vrecovery, \vballot^*, s)$, we have that
$\vrecovery=\TRUE$. Since the process $p_r$ will not move to the preceding epoch
until at least one process replies with $\PROBEACK$, this means that the probing
can never go beyond $\vballot$. Since the process $p_i$ is not a member of
$\vballot$, it cannot be included as a member of $s$ in $\vballot^*$, as
required.\qed

\subparagraph*{Proof sketch for Invariant~\ref{inv:main}.} We prove the
invariant by induction on $\vballot'$. Assume that the invariant holds for all
$\vballot' < \vballot''$. We now show it for $\vballot' = \vballot''$ by
induction on the length of the protocol execution. We only consider the most
interesting transition in line~\ref{alg:receivenewconfig-leader}, when a process
$p_i$ becomes a leader of $s$ at an epoch $\vballot''$. We show that after this
transition at $p_i$ we have $\prefix{\txn}{k} \prec \vtxn$,
$\prefix{\vote}{k} \prec \vvote$ and $\prefix{\payload}{k} \prec \vpayload$.

Since $p_i$ was chosen as the leader of $s$ at $\vballot''$, this process
replied with $\PROBEACK(\TRUE, \vballot'', s)$ to a
$\PROBE(\vballot'')$. Therefore, $p_i$ was a member of $s$ at an epoch
$\vballot^*<\vballot''$ that was being probed. Probing ends when at least one
process sends a $\PROBEACK(\TRUE, \vballot'', s)$. From
Invariant~\ref{inv:probe} and the assumption that all followers in $\vballot$
replied with $\ACCEPTACK$ to $\ACCEPT(\vballot, k, t, \pl, d)$, we can conclude
that probing could no have gone further than $\vballot$. Hence,
$\vballot\leq \vballot^*<\vballot''$.

Let $\vballot_0$ be the value of $\ballot[s]$ at $p_i$ right before the
transition at line~\ref{alg:receivenewconfig-leader}. We have
$\vballot_0\geq\vballot$, as otherwise $p_i$ would not be a member of $s$ at
$\vballot$ and by Invariant~\ref{inv:membership} could not be picked as the
leader of $s$ at $\vballot''$. It is also easy to show that
$\vballot_0< \vballot''$. Hence, $\vballot\leq\vballot_0<\vballot''$.

If $\vballot<\vballot_0$, then by the induction hypothesis, we have
$\prefix{\txn}{k} \prec \vtxn$, $\prefix{\vote}{k} \prec \vvote$ and
$\prefix{\payload}{k} \prec \vpayload$ right after the transition in
line~\ref{alg:receivenewconfig-leader}, as required. Assume now that
$\vballot_0=\vballot$. If $p_i$ was the leader of $s$ at $\vballot$, then we
trivially have $\prefix{\txn}{k} \prec \vtxn$, $\prefix{\vote}{k} \prec \vvote$
and $\prefix{\payload}{k} \prec \vpayload$ right after the transition in
line~\ref{alg:receivenewconfig-leader}, as required. Otherwise, by
Invariant~\ref{inv:probe}, $p_i$ must have received
$\ACCEPT(\vballot, k, t, \pl, d)$ and responded to it with
$\ACCEPTACK(s, \vballot, t)$ before the transition in
line~\ref{alg:receivenewconfig-leader}. Then the required follows from
Invariant~\ref{inv:synchrony}.\qed

\smallskip
\smallskip
\smallskip

We next state liveness properties of our protocol (we again defer proofs
to~\tr{app:proof-conc}). The reconfiguration procedure in the protocol will get
stuck if it cannot find an initialized process, which may happen if enough
processes crash, so that all shard data is lost. We now state conditions under
which this cannot happen. We associate two events with each configuration $e$ of
a shard $s$: \emph{introduction} and \emph{activation}. Introduction indicates
that the configuration comes into existence and is triggered when the
configuration is successfully persisted in the \external{}
(line~\ref{line:writezk}). Activation indicates that the configuration becomes
operational and is triggered when all the followers of the configuration have
processed the $\REPLICATELOGS$ messages sent by its leader
(line~\ref{alg:newview}).


Once a configuration has been activated, we say that it is \emph{active}. We
define its \emph{lifetime} as the time interval between its introduction and
when a succeeding configuration becomes active. Note that not every introduced
configuration necessarily becomes active, since its leader may never complete
the data transfer to the followers. To ensure our protocol is live we make the
following assumption, similar to the ones made by other protocols with changing
membership~\cite{ken-book, spiegelman2017dynamic}.

\begin{liveness}
\label{liveness-condition}
At least one member in each configuration is non-faulty throughout the lifetime
of a configuration.
\end{liveness}

The following two theorems show that, under this assumption, a single
reconfiguration makes progress.
\begin{theorem}\label{thm:livenessreconfiguration}
  If a process $p_r$ attempts to reconfigure a shard $s$ and no other process
  attempts to reconfigure $s$ simultaneously, then if $p_r$ is non-faulty for long
  enough, it will eventually introduce a new configuration.
\end{theorem}

\begin{theorem}
  If a configuration of a shard $s$ is introduced by a process $p_r$, then it will eventually be
  activated, provided no process
  attempts to reconfigure $s$ simultaneously, and $p_r$ and all the
  members of the configuration are non-faulty for long enough.
\end{theorem}

Finally, the following theorem shows that in the absence of failures or
reconfigurations, transaction certification makes progress.

\begin{theorem}
  Assume that the current configuration of each shard is active, all processes
  are aware of the current configuration of each shard, and no reconfiguration
  is in progress. If a transaction is submitted for certification, then it will
  eventually be decided, provided no reconfiguration is attempted and all the
  processes belonging to the current configuration of each shard are non-faulty
  for long enough.
\end{theorem}


\section{Exploiting RDMA}
\label{sec:rdmadiscussion}

\begin{figure*}[t]
\center
\includegraphics[width=\textwidth]{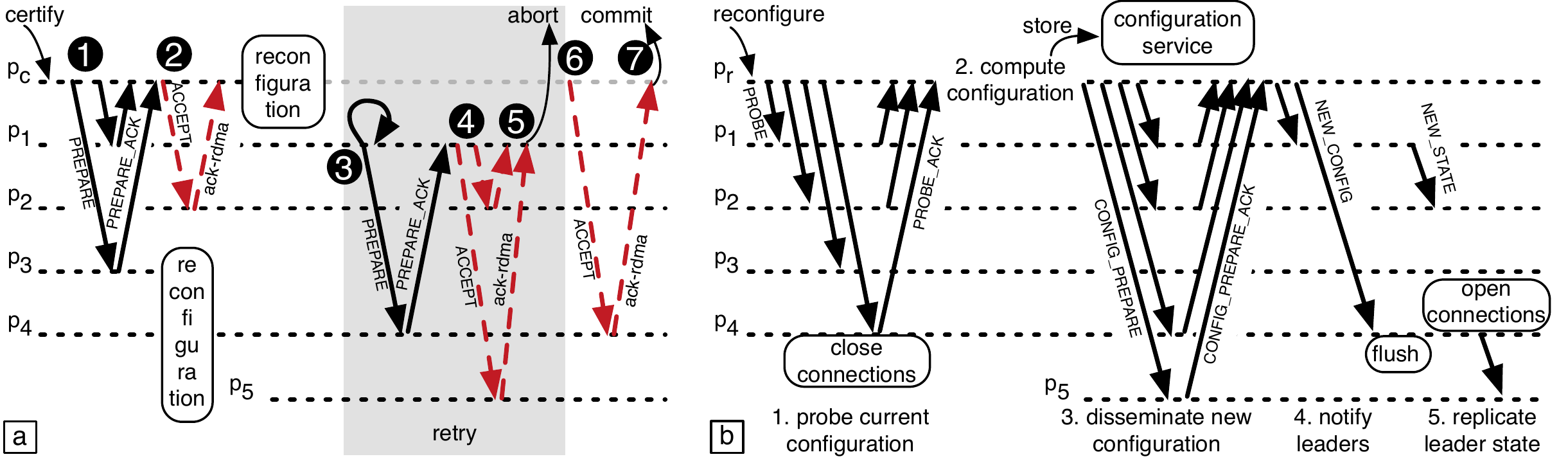}
\vspace{-10pt}
\caption{\rm Illustrations of (a) a counter-example showing the need to change
  the reconfiguration protocol when using RDMA; (b) the changed reconfiguration
  protocol. Red dashed lines denote RDMA operations. In both cases $p_1$ is the
  leader of shard $s_1$ and $p_2$ its follower; $p_3$ is the initial leader of
  shard $s_2$ and $p_4$ its follower. After the reconfiguration of $s_2$,
  process $p_4$ is the new leader and a fresh process $p_5$ is its follower.}
\label{fig:flowrdma}
\end{figure*}

We now present a variant of our protocol that uses Remote Direct Memory Access
(RDMA), which follows the design of the FARM system~\cite{farm,farm2}. By
comparing this protocol with that of \S\ref{sec:cft} we highlight the trade-offs
required by the use of RDMA. Due to space constraints, we defer the pseudocode
of our protocol to~\tr{sec:rdma} and describe the required changes in the protocol
of \S\ref{sec:cft} only informally.

We assume the same system model as in \S\ref{sec:tcs}, except that processes can
communicate using RDMA. This allows a machine to access the memory of another
machine over the network without involving the latter's CPU, thus lowering
latency. Like FARM, our protocol uses RDMA to implement a primitive for
point-to-point communication between processes with the following interface. The
primitive allows a sender process to reliably send a message $m$ to a receiver
process $p_j$ (\sendrdma($m, p_j$)) by remotely writing into a specific memory
region of $p_j$. The sender then gets an acknowledgement when the message
reaches the receiver's memory (\ackrdma($m, p_j$)), sent by the receiver's
network interface card (NIC) without interrupting its CPU. The receiver is
notified at a later point that a new message is available
(\deliverrdma($m, p_j$)). Hence, the guarantee provided by \ackrdma($m, p_j$) is
that the receiver will eventually deliver the message $m$, even if the sender
crashes, since the message is already in the receiver's memory.
The operation \openconnection($p_i$) grants $p_i$ access to a region of the
caller's memory, and \closeconnection($p_i$) revokes it. Once the latter
operation completes, $p_i$ cannot send any message to the caller using
\sendrdma. Finally, we assume that the communication primitive includes another
operation: \emptybuffers. This operation blocks the caller until it has
delivered all messages addressed to it that have been acknowledged by its NIC
through an \ackrdma.

To implement the above primitive, the receiver usually keeps a circular buffer
in memory for each process that may send it a message~\cite{farm-first,
  rdma-mpi}. 
The operation \sendrdma($m, p_j$) issued by a process $p_i$ appends a message to
the corresponding buffer at the receiver using RDMA writes. Receivers
periodically pull messages from the buffers and deliver them to the application
via \deliverrdma. If a buffer at a process $p_j$ gets full, the associated
sender process will not be able to send a message to $p_j$ until the latter
pulls some messages.

Following FARM, we use the above RDMA-based communication primitive in our
protocol to persists votes and decisions (steps 2 and 3 of
Figure~\ref{fig:flow}a). This requires the following changes to the protocol in
Figure~\ref{fig:protocol}. First, $\ACCEPT$ and $\DECISION$ messages are sent
using \sendrdma instead of \Send{} (lines~\ref{alg:send-accept}
and~\ref{alg:send-decision-groups2}). Second, the followers do not send explicit
$\ACCEPTACK$ messages to transaction coordinators (line~\ref{alg:accept-ack});
instead, the latter act once they receive an RDMA acknowledgement \ackrdma. This
makes the checks at lines~\ref{alg:check-accept} and~\ref{alg:check-decision}
redundant, as followers cannot reject $\ACCEPT$ or $\DECISION$ messages under
any circumstance. The practical rationale for these changes is that persisting a
transaction $t$ at followers using RDMA minimizes the time during which the
transaction is prepared at leaders, which requires them to vote abort on all
transactions conflicting with $t$ (via the certification function $g_s$,
\S\ref{sec:tcs}); this results in lower abort
rates~\cite{farm,binnig}. Transaction processing at followers (e.g., adding them
to the local copy of the certification order, line~\ref{alg:receive-accept}) is
done off the critical path of certification.

Unfortunately, the above changes to the failure-free path of the protocol do not
preserve correctness without changes to reconfiguration, as illustrated by an
example execution in Figure~\ref{fig:flowrdma}a. In this execution, two shards
$s_1$ and $s_2$ are involved in the certification of a transaction $t$,
coordinated by a process $p_c$ from a third shard. The transaction is prepared
to commit at the leaders $p_1$ and $p_3$ of both shards (step \circled{1}), and
the {\sc commit} vote from the leader of $s_1$ ($p_1$) is persisted at the
follower $p_2$ using RDMA (step \circled{2}). Before the coordinator $p_c$
persists the vote from the leader $p_3$ of $s_2$ at the follower $p_4$, the
leader $p_3$ is suspected of failure and a reconfiguration is triggered at shard
$s_2$. This promotes the follower $p_4$ to a new leader and brings online a
fresh follower $p_5$. Next, the leader $p_1$ of $s_1$ suspects the coordinator
$p_c$ of failure and triggers a reconfiguration to remove it. Once $p_c$ is
removed from its shard, $p_1$ retries the processing of $t$ (step \circled{3},
line~\ref{alg:retry} in Figure~\ref{fig:protocol}). The new leader $p_4$ of
$s_2$ does not know about $t$, so this results in the transaction being aborted,
because its payload at shard $s_2$ is thought to be lost (steps \circled{4} and
\circled{5}). But now the coordinator $p_c$, who did not actually fail and still
believes $s_2$ is in the old configuration, finishes its processing by
persisting the {\sc commit} vote of the old leader $p_3$ of $s_2$ at the old
follower $p_4$, which is now the new leader of $s_2$ (step \circled{6}). Since
this is done via RDMA, $p_4$ cannot reject the vote and, thus, $p_c$ commits the
transaction (step \circled{7}). This violates safety, as two contradictory
results have been externalized. The protocol in \S\ref{sec:cft} is not subject
to this problem, because in that protocol the new leader $p_4$ of the shard
$s_2$ would reject the $\ACCEPT$ message due to the failure of the check at
line~\ref{alg:check-accept}.


To make the RDMA-based protocol correct, we need to change the reconfiguration
protocol so that {\em the whole system} participates in reconfiguration instead
of just the affected shard. Figure~\ref{fig:flowrdma}b illustrates the message flow of the redesigned
reconfiguration protocol. Processes now maintain a single epoch variable instead of a vector. The
data structures maintained by the external \external and its interface are
adjusted accordingly. Like in our previous commit protocol, the process
$p_r$ performing reconfiguration first probes previous configurations by sending
$\PROBE$ messages. However, $p_r$ now probes {\em all} shards.
A process receiving $\PROBE$ handles it as before (line~\ref{alg:probe}), but
additionally closes all incoming RDMA connections using \closeconnection, which
guarantees that the process accepts no more transactions at its previous
epoch. This is needed because, due to communication via RDMA, the protocol
cannot longer leverage the safety check at line~\ref{alg:check-accept}. The
logic of the reconfiguring process is also changed: after this process computes
the new configuration and stores it in the \external{}
(line~\ref{line:writezk}), the process sends a new $\CONFIGPREPARE$ message to
{\em all} processes in the configuration. Upon receiving $\CONFIGPREPARE$, a
process updates its locally stored configuration and replies with a
$\CONFIGPREPAREACK$ message. This ensures that the whole system is aware of the
new configuration before it is activated. Only after this does the reconfiguring
process send a $\NEWCONFIG$ message to the leaders of the new
configuration. Upon receiving $\NEWCONFIG$
(line~\ref{alg:receivenewconfig-leader}), a leader $p_l$ first calls
\emptybuffers. This guarantees that all the messages that have been acknowledged
as having reached $p_l$'s memory will be replicated to followers in
$\REPLICATELOGS$ messages; this is necessary since transaction coordinators may
have already externalized decisions taken based on these
acknowledgements. Finally, processes open RDMA connections to all other
processes in the configuration using \openconnection: a leader after sending
$\REPLICATELOGS$ to its followers, and followers upon receiving $\REPLICATELOGS$
(line~\ref{alg:newview}).

The new protocol guarantees that: (*) if a process receives an $\ACCEPT$ message
for a transaction $t$ while at epoch $\vballot$, then the leader that prepared
$t$ was at epoch $\vballot$ when it prepared this transaction. This property is
key in proving the correctness of the protocol, as it provides the same
guarantees as the removed guard in line~\ref{alg:check-accept}, which we could
not leverage due to the use of RDMA. The property (*) holds because: {\em (i)}
at any time, a process only maintain RDMA connections to the members of its
current epoch; and {\em (ii)} before persisting a vote at a follower, the
coordinator of a transaction checks that the transaction was prepared in its
current epoch (line~\ref{alg:coord-check}).

We now show how the revised reconfiguration protocol prevents the bug in
Figure~\ref{fig:flowrdma}a. In this protocol, when $p_c$ attempts to persist the
{\sc commit} vote at $p_4$ (step \circled{6}), the latter will be already aware
that $p_c$ has been removed from the system and will close the RDMA
connection to it. Thus, $p_c$ will be unable to persist the vote at $p_4$ (this would
violate the property (*)) and will never gather enough acknowledgements to
decide the transaction. Hence, no contradictory results will be externalized. We
state and prove the correctness of the RDMA-based protocol
in~\tr{sec:rdma}.


\section{Related Work and Discussion}
\label{sec:discussion}

Our protocols are inspired by the recent FARM system for transaction processing,
which also uses $f+1$ replicas per shard and deals with failures using
reconfiguration~\cite{farm,farm2}. FARM was presented as a complete database
system with a number of optimizations, including the use of RDMA. In contrast,
our work distills the core ideas of FARM into protocols solving the well-defined
transaction certification problem, parametric in the isolation level provided
and rigorously proven correct. This allows us to simplify some aspects of the
FARM design. In particular, FARM has a more complex way of determining the state
of the new leader upon a reconfiguration, which merges the states from all
surviving replicas of the previous configuration. In contrast, our protocols
take the state of any single initialized replica. Our reconfiguration protocols
also provide better fault-tolerance guarantees on a par with those of existing
ones~\cite{ken-book,spiegelman2017dynamic}. This is because, like Vertical Paxos
I~\cite{vertical-paxos}, our protocols look through a sequence of configurations
to find the new leader, whereas FARM only considers the previous
configuration. Hence, FARM reconfiguration can get stuck even when there exists
a non-faulty replica with the necessary data. Finally, by presenting two related
protocols using message passing and RDMA, we are able to identify the price of
exploiting RDMA---having to reconfigure the whole system instead of a single
shard.

There have been a number of protocols for solving the atomic commit problem,
which requires reaching a decision on a single
transaction~\cite{2pc,Hadzilacos1990,nbac,dwork-skeen}. In contrast to these
works, our protocol solves the more general problem of implementing a
Transaction Certification Service, which requires reaching decisions on a stream
of transactions. This problem more faithfully reflects the requirements of
modern transaction processing systems~\cite{discpaper}.

Our protocol weaves together two-phase commit (2PC)~\cite{2pc} and Vertical
Paxos~\cite{vertical-paxos}, instead of using Paxos replication as a black
box. This is similar to several existing sharded systems for transaction
processing, which integrate protocols for distribution and
replication~\cite{uw-inconsistent,mdcc,replicated-commit,discpaper}. However,
these systems considered a static set of $2f+1$ processes per shard, whereas we
assume $f+1$ processes and allow the system to be reconfigured. Achieving this
correctly is nontrivial and requires a subtle interplay between the
reconfigurable replication mechanism and cross-shard coordination. For example,
as we explained in \S\ref{sec:cft}, on failures our protocol may lose
information about transactions that influenced votes on other transactions, but
this does not violate correctness. As is well-known~\cite{cheappaxos}, using
$f+1$ instead of $2f+1$ replicas results in somewhat weaker availability
guarantees: upon a single failure, our protocols have to stop processing
transactions while the system is reconfigured.


\subparagraph*{Acknowledgments.} We thank Dushyanth Narayanan for discussions
about FARM. This research was supported by an ERC grant RACCOON.

\bibliographystyle{abbrv}
\bibliography{biblio}

\iflong
\appendix
\clearpage

\newpage
\section{Correctness of the Protocol}
\label{app:proof-conc}

Figure~\ref{fig:inv-rest} summarizes additional invariants that, together with the
invariants listed in Figure~\ref{fig:inv}, are used to prove the correctness of the protocol.
We first prove the nontrivial
Invariants~\ref{inv:synchrony},~\ref{inv:probe},~\ref{inv:accepted},~\ref{inv:committed}
and~\ref{inv:consensus} that were not proved in \S\ref{sec:correctness}. We then prove Theorem~\ref{thm:conc}.

\subsection{Proof of Invariants}

\begin{figure}[t]
\begin{enumerate}[leftmargin=15pt, label={\arabic*.}, ref={\arabic*}]
\setcounter{enumi}{5}
\item\label{inv:unique-accept} If $\ACCEPT(\vballot, k, t_1, \pl_1, d_1)$ and
  $\ACCEPT(\vballot, k, t_2,  \pl_2, d_2)$ messages are sent to the same shard, then
  $t_1 = t_2$, $\pl_1 = \pl_2$ and $d_1 = d_2$.

\item\label{inv:payload} 
\begin{enumerate}[leftmargin=0.6cm]
  \item 
If at a process in a shard $s$ we have $\vote[k]=\COMMIT$ then
    $\payload[k]=(\pl\mid s)$, where $\pl$ is the
  payload of $\txn[k]$ submitted in the $\Certify(\txn[k],\pl)$.
\item If at a process in a shard $s$ we have $\vote[k]=\ABORT$ then
    $\payload[k]=\{(\pl\mid s)\mid \varepsilon\}$, where $\pl$ is the
  payload of $\txn[k]$ submitted in the $\Certify(\txn[k],\pl)$.
  \end{enumerate}

\item\label{inv:cballot} At any process, we always have $\cballot\geq\ballot[s_0]$.

\item\label{inv:accept-tx} If $\ACCEPT(\vballot, k_1, t, \_, \_)$ and
  $\ACCEPT(\vballot, k_2, t, \_, \_)$ messages are sent to the same shard, then
  $k_1 = k_2$.

\item\label{inv:unique-tx} At any process, all transactions in the $\txn$ array
  are distinct.

\item\label{inv:accepted} 
  \begin{enumerate}[leftmargin=0.6cm]
  \item Assume that all followers in $s$ at $\vballot_1$ have received $\ACCEPT(\vballot_1, k, t_1, \pl_1,
    d_1)$ and replied with
    $\ACCEPTACK(s, \vballot_1, k, t_1, d_1)$. Assume that all followers in $s$
    at $\vballot_2$ have received $\ACCEPT(\vballot_2, k, t_2, \pl_2,
    d_2)$ and replied with
    $\ACCEPTACK(s, \vballot_2, k, t_2, d_2)$. Then $t_1 = t_2$, $\pl_1 = \pl_2$ and $d_1 = d_2$.
  \item Assume that all followers in $s$ at $\vballot_1$ have received $\ACCEPT(\vballot_1, k_1, t, \pl_1,
    d_1)$ and replied with
    $\ACCEPTACK(s, \vballot_1, k_1, t, d_1)$. Assume that all followers in $s$ at
    $\vballot_2$ have received $\ACCEPT(\vballot_2, k_2, t, \pl_2,
    d_2)$ and replied with
    $\ACCEPTACK(s, \vballot_2, k_2, t, d_2)$. Then $k_1 = k_2$, $\pl_1 = \pl_2$ and
    $d_1 = d_2$.
\end{enumerate}

\item\label{inv:committed}
\begin{enumerate}[leftmargin=0.6cm]
  \item If at a process in a shard $s$ we have $\ballot[s] = \vballot'$,
$\phase[k] = \DECIDED$ and $\decision[k] = d$, then a $\DECISION(\vballot, k, d)$
message has been sent to $s$, where $\vballot \le \vballot'$.
\item If at a process we have $\phase[k] = \DECIDED$
  and $\decision[k] = \COMMIT$, then $\vote[k] = \COMMIT$.
  \end{enumerate}
\end{enumerate}
\caption{\rm Additional invariants of the atomic commit protocol used in its
  proof of correctness.}
\label{fig:inv-rest}
\end{figure}

\subparagraph*{Proof of Invariant~\ref{inv:synchrony}.} Assume that a
process $p_i$ in $s$ at $\vballot$ receives $\ACCEPT(\vballot, k, t, \pl,
d)$ and replies with $\ACCEPTACK$. We prove that, after the transition
and while $\ballot[s]=\vballot$, $p_i$ has $\prefix{\txn}{k} \prec
\prefix{\vtxn}{k}$, $\prefix{\vote}{k} \prec
\prefix{\vvote}{k}$ and $\prefix{\payload}{k} \prec
\prefix{\vpayload}{k}$, where $\vtxn$, $\vvote$ and $\vpayload$ are the values of the
arrays $\txn$, $\vote$ and $\payload$ at the leader $p_l$ of $s$ at $\vballot$ when it
sent the corresponding message $\PREPAREACK(\vballot, s, k, t, \pl, d)$. 

If $p_i$ processes $\ACCEPT(\vballot, k, t, \pl,
d)$, then $p_i$ has $\ballot=\vballot$. Thus, $p_i$ has processed $\REPLICATELOGS(\vballot,
\vtxn',\vpayload',\vvote',\_,\_)$ before. After processing this message, $p_i$ has $\txn=\vtxn'$,
$\vote=\vvote'$ and $\payload=\vpayload'$ where $\vtxn'$, $\vvote'$ and $\vpayload'$ are the values of the
arrays $\txn$, $\vote$ and $\payload$ at the leader $p_l$ of $s$ at
$\vballot$ when it sent the $\REPLICATELOGS$ message. Let
$k'=\length(\vtxn')$. By lines~\ref{line:unique_accept},~\ref{line:next} and ~\ref{line:setnext} we have that $\prefix{\vtxn}{k'} =
\vtxn'$, $\prefix{\vvote}{k'} = \vvote'$ and $\prefix{\vpayload}{k'} = \vpayload'$. Therefore, $p_i$ has $\prefix{\txn}{k'} \prec \prefix{\vtxn}{k'}$,
$\prefix{\vote}{k'} \prec \prefix{\vvote}{k'}$ and
$\prefix{\payload}{k'} \prec \prefix{\vpayload}{k'}$ while
$\ballot[s]=\vballot$. Furthermore, after processing $\ACCEPT(k, t, \pl,
d)$, $p_i$ has $\txn[k]=\vtxn[k]$, $\vote[k]=\vvote[k]$ and
$\payload[k]=\vpayload[k]$. 
By Invariant~\ref{inv:accept-tx}, $p_i$ has $\txn[k]=\vtxn[k]$, $\vote[k]=\vvote[k]$ and
$\payload[k]=\vpayload[k]$ while $\ballot[s]=\vballot$. 

We now prove that after processing $\REPLICATELOGS(\vballot,
\vtxn',\vpayload',\vvote',\_,\_)$ and while
$\ballot[s]=\vballot$, $p_i$ has $\txn[k'']\in\{\vtxn[k''], \bot\}$,
$\vote[k'']\in\{\vvote[k''], \bot\}$ and
$\payload[k'']\in\{\vpayload[k''], \bot\}$ for any $k''$ such that
$k'<k''<k$. We prove it by induction on
the length of the protocol execution from the moment in which $p_i$ has
processed $\REPLICATELOGS(\vballot,
\vtxn',\vpayload',\vvote',\_,\_)$. The validity of the
property can be affected by only the transition at
line~\ref{alg:receive-accept}. Let $\ACCEPT(\vballot, k^*, t^*, \pl^*,
d^*)$ be the message that triggers the transition. Assume that
$k'<k^*<k$, as otherwise the transition does not affect the validity
of the property. By the induction hypothesis, $p_i$ has $\txn[k'']\in\{\vtxn[k''], \bot\}$,
$\vote[k'']\in\{\vvote[k''], \bot\}$ and
$\payload[k'']\in\{\vpayload[k''], \bot\}$ for any $k''\neq
k^*$ such that $k'<k''<k$ after processing $\ACCEPT(\vballot, k^*, t^*, \pl^*,
d^*)$. Also, after processing $\ACCEPT(\vballot, k^*, t^*, \pl^*,
d^*)$, $p_i$ has $\txn[k^*]=t^*$, $\vote[k^*]=d^*$ and
$\payload[k^*]=\pl^*$. By lines~\ref{line:unique_accept},~\ref{line:next} and ~\ref{line:setnext}, $\vtxn[k^*]=t^*$, $\vvote[k^*]=d^*$ and
$\vpayload[k^*]=\pl^*$. Then, after processing $\ACCEPT(\vballot, k^*, t^*, \pl^*,
d^*)$, $p_i$ has $\txn[k^*]=\vtxn[k^*]$, $\vote[k^*]=\vvote[k^*]$ and
$\payload[k^*]=\vpayload[k^*]$. This proves that, after processing $\REPLICATELOGS(\vballot,
\vtxn',\vpayload',\vvote',\_,\_)$ and while
$\ballot[s]=\vballot$, $p_i$ has $\txn[k'']\in\{\vtxn[k''], \bot\}$,
$\vote[k'']\in\{\vvote[k''], \bot\}$ and
$\payload[k'']\in\{\vpayload[k''], \bot\}$ for any $k''$ such that
$k'<k''<k$. We have already proved that (i) $p_i$ has $\prefix{\txn}{k'} \prec \prefix{\vtxn}{k'}$,
$\prefix{\vote}{k'} \prec \prefix{\vvote}{k'}$ and $\prefix{\payload}{k'}
\prec \prefix{\vpayload}{k'}$ after processing $\REPLICATELOGS(\vballot,
\vtxn',\vpayload',\vvote',\_,\_)$ and while
$\ballot[s]=\vballot$; and that (ii) $p_i$ has $\txn[k]=\vtxn[k]$, $\vote[k]=\vvote[k]$ and
$\payload[k]=\vpayload[k]$ after processing $\ACCEPT(\vballot, k, t, \pl,
d)$ and while $\ballot[s]=\vballot$. Hence, $p_i$ has $\prefix{\txn}{k} \prec
\prefix{\vtxn}{k}$, $\prefix{\vote}{k} \prec
\prefix{\vvote}{k}$ and $\prefix{\payload}{k} \prec
\prefix{\vpayload}{k}$ after processing $\ACCEPT(\vballot, k, t, \pl,
d)$ and while $\ballot[s]=\vballot$, as required. \qed

\subparagraph*{Proof of Invariant~\ref{inv:probe}.} When $p_i$ processed $\PROBE(\vballot)$, it set
$\status=\RECOVERING$. This prevents $p_i$ from processing any $\ACCEPT$ message
until it processes a $\NEWCONFIG(\vballot^*, \_)$ or a
$\REPLICATELOGS(\vballot^*, \_,\_,\_,\_,\_,\_)$. When $p_i$
processed $\PROBE(\vballot)$, it also sets
$\cballot=\vballot$. By the checks in lines~\ref{alg:receivenewconfig-leader}
  and~\ref{alg:check-replicatelogs} and by the fact that $\cballot$
  can never decrease, this guarantees that $p_i$ only
handles any of these messages if $\vballot^*\geq\cballot$. Hence,
by the time $p_i$ is able to process $\ACCEPT$ messages again it will have
$\ballot[s]=\vballot^*>\vballot'$. By the check in
line~\ref{alg:check-accept} and the fact that the protocol trivially
guarantees that $\ballot[s]$ never decreases, $p_i$
will never send $\ACCEPTACK(s, \vballot', \_, \_, \_)$ after sending a
$\PROBEACK(\_, \vballot, s)$, as required. \qed

\subparagraph*{Proof of Invariant~\ref{inv:accepted}.} \textit{(a)}
Assume that all followers in $s$ at $\vballot_1$ have received
$\ACCEPT(\vballot_1, k, t_1, \pl_1, d_1)$ and replied with
$\ACCEPTACK(s, \vballot_1, k, t_1, d_1)$. Assume that all followers in $s$ at
$\vballot_2$ have received $\ACCEPT(\vballot_2, k, t_2, \pl_2, d_2)$
and replied with $\ACCEPTACK(s, \vballot_2, k, t_2, d_2)$. Assume without loss of
generality that $\vballot_1 \le \vballot_2$. If $\vballot_1 = \vballot_2$, then by
Invariant~\ref{inv:unique-accept} we must have $t_1 = t_2$, $\pl_1 = \pl_2$ and $d_1 =
d_2$. Assume now that $\vballot_1 < \vballot_2$. By
Invariant~\ref{inv:main}, when the leader of $s$ at $\vballot_2$ sent the
$\PREPAREACK(\vballot_2, s, k, t_2, \pl_2, d_2)$ message it has
$\txn[k] = t_1$, $\vote[k] = d_1$ and $\payload[k] = \pl_1$. But then due
to the check at line~\ref{line:unique_accept}, we again must have $t_1
= t_2$, $\pl_1 = \pl_2$ and $d_1 = d_2$.

{\em (b)} Assume that all followers in $s$ at $\vballot_1$ have
received $\ACCEPT(\vballot_1, k_1, t, \pl_1, d_1)$ and replied with
$\ACCEPTACK(s, \vballot_1, k_1, t, d_1)$. Assume that all followers in $s$ at
$\vballot_2$ have received $\ACCEPT(\vballot_2, k_2, t, \pl_2, d_2)$
and replied with $\ACCEPTACK(s, \vballot_2, k_2, t, d_2)$. Assume without loss of
generality that $\vballot_1 \le \vballot_2$. We first
show that $k_1 = k_2$. If $\vballot_1 = \vballot_2$, then we must have $k_1 = k_2$ by 
Invariant~\ref{inv:accept-tx}. Assume now that $\vballot_1 < \vballot_2$. By
Invariant~\ref{inv:main}, when the leader of $s$ at $\vballot_2$ sent the
$\PREPAREACK(\vballot_2, s, k_2, t, \pl_2, d_2)$ message it
has $\txn[k_1] = t$. But then due to the check at line~\ref{line:unique_accept}
and Invariant~\ref{inv:unique-tx}, we again must have $k_1 = k_2$. Hence,
$k_1 = k_2$. But then by Invariant~\ref{inv:accepted}a we must also have
$\pl_1 = \pl_2$ and $d_1 = d_2$. \qed

\subparagraph*{Proof of Invariant~\ref{inv:committed}.}  \textit{(a)}
Assume that a process $p_i$ in shard $s$ has $\ballot[s]=\vballot'$,
$\phase[k]=\DECIDED$ and $\decision[k] = d$. We show that then a $\DECISION(\vballot, k, d)$
message has been sent to $s$, where $\vballot \le \vballot'$. We prove the invariant by
induction on the length of the protocol execution. The validity of the
property can be affected by only the transitions at
lines~\ref{alg:receive-decision} and~\ref{alg:newview}. First, consider the transition
at line~\ref{alg:receive-decision}. By the induction hypothesis, $p_i$
satisfies the property before handling the $\DECISION(\vballot, k, d)$ message that
causes the transition. Given that the message is only handled if
$\vballot\leq \vballot'$, lines~\ref{alg:decision-store}
and~\ref{alg:decision-phase} trivially preserve the invariant. Finally, consider the transition
at line~\ref{alg:newview}. The transition is triggered when $p_i$ receives a
$\REPLICATELOGS(\vballot', \_, \_, \_, \_, \vdecision, \vphase)$.  By the induction
hypothesis, the leader of $s$ at $\vballot'$ satisfies the required
before the transition. The process $p_i$ simply substitutes its $\decision$ and
$\phase$ arrays by the arrays $\vdecision$ and
$\vphase$. Therefore, $p_i$ will also satisfy the required
after the transition.

\textit{(b)} Follows from item (a) and Invariant~\ref{inv:main}.\qed

\subparagraph*{Proof of Invariant~\ref{inv:consensus}.} Follows from
Invariant~\ref{inv:accepted}, since, if a coordinator has computed the
final decision on a transaction, then all followers in each relevant
shard at a given epoch have accepted a corresponding vote.\qed

\subsection{Proof of Theorem~\ref{thm:conc}}

To facilitate the proof of Theorem~\ref{thm:conc}, we first introduce a
low-level specification \lltcsone{}, and prove that it is correctly implemented
by the atomic commit protocol (Lemma~\ref{thm:constraints}). We then show that
every history satisfying \lltcsone{} is correct with respect to $f$
(Lemma~\ref{thm:2pc}).  The low-level specification \lltcsone{} is defined as
follows.

\begin{quote}
  Consider a history $\hist$. Let $T$ denote the set of transactions $t$ such
  that $\Certify(t,\_)$ is an event in $\hist$, and $d[t]$ denote the decision
  value $d$ of $t\in T$ if $\Decide(t, d)$ is an event in $\hist$. The history
  $\hist$ satisfies \lltcsone{} if for some of transactions $t\in T$ and shards
  $s \in \shards(t)$ there exist $d_s[t] \in \D$, $\pos_s[t]\in
  \mathbb{N}$, $\pload_s[t]\in \Payload$ and
  $T_s[t], P_s[t] \in 2^{\Txn}$ such that all the constraints in
  Figure~\ref{fig:constraints} are satisfied.  A protocol is a correct
  implementation of \lltcsone{} if each of its finite histories satisfies
  \lltcsone{}.
\end{quote}

\begin{figure}[t]
\begin{fleqn}
\begin{gather}
\label{decision}
\forall t.\, d[t] = \bigsqcap\{d_s[t] \mid s \in \shards(t)\}
\end{gather}
\\
\begin{gather}
\label{unique-pos}
\forall t_1, t_2, s.\, t_1 \not= t_2 \implies \pos_s[t_1] \not=
\pos_s[t_2]
\end{gather}
\\
\begin{multline}
\label{property-payload}
\forall t,l,s.\,\Certify(t, \pl)\in h \implies\\ 
\hspace{2.1cm} (d_s[t] =
\COMMIT\implies \pload_s[t] = (\pl\mid s)) \wedge {}\\
(d_s[t] =
\ABORT\implies \pload_s[t]\in \{(\pl\mid s), \varepsilon\})
\end{multline}
\\
\begin{multline}
\label{compute1}
\forall t, s.\, d_s[t] \sqsubseteq \\f_s(\pload_s(T_s[t]),
\pload_s[t]) \sqcap g_s(\pload_s(P_s[t]), \pload_s[t])
\end{multline}
\\
\begin{multline}
\label{compute2}
\forall t, s.\, T_s[t] = \\\{t' \mid \pos_s[t'] < \pos_s[t] \wedge d[t'] = \COMMIT\}
\setminus P_s[t]
\end{multline}
\\
\begin{gather}
\label{compute4}
\forall t, s.\, P_s[t] \subseteq \{t' \mid \pos_s[t'] < \pos_s[t] \wedge d_s[t'] = \COMMIT\}
\end{gather}
\\
\begin{multline}
\label{rt}
\forall t, t', s.\, t' \rt t \wedge 
s \in \shards(t') \cap \shards(t) {\implies} \\\pos_s[t'] < \pos_s[t]
\end{multline}
\\
\begin{gather}
\label{acyclic}
{\rt} \cup {\decrel} \mbox{ is acyclic},
\end{gather}
\\
where
\\
\begin{multline*}
\forall x,y\in\{\ABORT\mid\COMMIT\}.\,x\sqsubseteq y {\iff} \\x=y \vee (x=\ABORT\wedge y=\COMMIT)
\end{multline*}
\\
\begin{equation*}
\forall t, t'.\, t' \rt t {\iff} \Decide(t', \_) \prec_\hist
\Certify(t, \_)
\end{equation*}
\\
\begin{multline*}
\forall t, t'.\, t' \decrel t {\iff} \exists s.\, t' \in T_s[t] \vee {}\\
(\pos_s[t'] \,{<}\, \pos_s[t] \wedge d_s[t'] \,{=}\, \COMMIT \wedge
d[t'] \,{=}\, \ABORT \wedge t' \,{\not\in}\, P_s[t])
\end{multline*}
\end{fleqn}
\caption{\rm Constraints on the votes computed by the atomic
  commit protocol. In (\ref{compute1}), we lift the array $\pload_s$ to a set of transactions.}
\label{fig:constraints}
\end{figure}

\begin{lemma}\label{thm:constraints}
  The atomic commit protocol in
  Figures~\ref{fig:protocol} is a correct implementation
  of  \lltcsone{}. 
\end{lemma}

\subparagraph*{Proof}  Fix a finite execution of the
atomic commit protocol with a history $\hist$. Let $T$ be
the set of transactions $t$ such that $\Certify(t, \pl)$ occurs in $h$. For some of
transactions $t \in T$, $\pl\in\Payload$, and shards $s \in \shards(t)$, we define the
certification order position $\pos_s[t]$, 
$\pload_s[t]$ and a vote $d_s[t]$ computed by the
protocol as follows:
\begin{quote}
  Consider $t \in T$ and $s \in \shards(t)$. Assume that all followers
  in $s$ at $\vballot$ received $\ACCEPT(\vballot, k, t, \pl, d)$ and responded to it with
  $\ACCEPTACK(s, \vballot, k, t, d)$. 
Then, we let $\pos_s[\txn[k]] = k$, $\pload_s[\txn[k]] = \pl$ and $d_s[\txn[k]] = d$.
\end{quote}
According to Invariants~\ref{inv:main} and~\ref{inv:unique-tx}, this defines
$\pos_s[t]$, $\pload_s[t]$ and $d_s[t]$ uniquely and~(\ref{unique-pos}) in
Figure~\ref{fig:constraints} holds. Furthermore, by the structure of the handler
at line~\ref{alg:receive-acceptack}, for each $t$ such that $\Decide(t, d[t])$
occurs in $h$, $d_s[t]$ is defined for all $s \in \shards(t)$
and~(\ref{decision}) holds. By Invariant~\ref{inv:payload}, (\ref{property-payload}) holds. 

We now prove (\ref{rt}). Consider $t, t', s$ such that
$$
\Decide(t, \_) \prec_\hist \Certify(t', \_) \wedge s \in \shards(t) \cap \shards(t').
$$
Let $\DECISION(\vballot, \pos_s[t], \_)$ be the message sent to the shard $s$ when the
$\Decide(t, \_)$ action was generated. Let $\vballot'$ be some epoch at which
$\pos_s[t']$ is defined according to the above definition. Assume
first that $\vballot' < \vballot$. Then by Invariant~\ref{inv:main} when the leader of $\vballot$ starts
operating, it has $\txn[\pos_s[t']] = t'$. But then
$\Certify(t', \_)$ must have
occurred before the $\Decide(t, \_)$. Hence, $\vballot \le \vballot'$. By Invariant~\ref{inv:main}
when the leader of $s$ at $\vballot'$ receives $\PREPARE(t', \_)$, it has $\txn[\pos_s[t]] =
t$. But then $\pos_s[t] < \pos_s[t']$, which proves~(\ref{rt}).

We prove~(\ref{compute1})-(\ref{compute4}) using the following
proposition. 

\begin{proposition}\label{prop:vote-compute}
  The following always holds at any process in a shard $s$:
\begin{equation}\label{define-process-t-p}
\begin{array}{@{}l@{}}
\forall k.\, (\vote[k] \mbox{ is defined}) \implies \\
\exists T, P.\, 
\vote[k] \sqsubseteq f_s(\fploadprocess(T, \txn, \payload),
  \payload[k]) \sqcap {}\\
\hspace{2.1cm} g_s(\fploadprocess(P, \txn, \payload), \payload[k]) \wedge {}
\ms
T = \{\txn[k'] \mid k' < k \wedge \vote[k'] = {\normalfont \COMMIT} \wedge {}\\
\hspace{.7cm} d[\txn[k']] = {\normalfont \COMMIT}\}
\setminus P \wedge {}
\ms
P \subseteq \{\txn[k'] \mid k' < k \wedge \vote[k'] = {\normalfont \COMMIT}\} \wedge {}
\ms
(\forall k'.\, \txn[k'] \in T {\implies} \\
\hspace{1.7cm} (\DECISION(\txn[k'], {\normalfont \COMMIT}) \mbox{ has been sent})) \wedge {}
\ms
(\forall k' < k.\, \vote[k'] = {\normalfont \COMMIT} \wedge \txn[k'] \not\in T \cup P
{\implies}\\
 \hspace{1.7cm} (\DECISION(\txn[k'], {\normalfont \ABORT}) \mbox{ has been sent}));
\end{array}
\end{equation}
where the function $\fploadprocess:(\Txn\times(\mathbb{N} \to
\Txn)\times(\mathbb{N} \to \Payload)) \to \Payload$ determines the
payload that a process has stored for a given transaction, i.e., for
any transaction $t\in\Txn$, and arrays $\vtxn\in \mathbb{N} \to
\Txn$ and $\vpayload\in \mathbb{N} \to \Payload$, $\fploadprocess(t,\vtxn,
\vpayload)=\{\vpayload[k]\mid t=\vtxn[k]\}$. We lift the function to sets
of transactions: for
any set of transactions $T\subseteq\Txn$ and arrays $\vtxn\in \mathbb{N} \to
\Txn$ and $\vpayload\in \mathbb{N} \to \Payload$, we have $\fploadprocess(T,\vtxn,
\vpayload)=\{\fploadprocess(t,\vtxn,\vpayload)\mid t\in T\}$.
\end{proposition}

\subparagraph*{Proof.} We prove this by induction on the length of the protocol
execution.  The validity of the above property can be nontrivially affected only
by the transitions at
lines~\ref{alg:compute-vote},~\ref{alg:forced-vote},~\ref{alg:accept-vote},
and~\ref{alg:new-vote}.

First consider the transition at line~\ref{alg:compute-vote}, which computes
$\vote[k]$ as follows:
$$
\begin{array}{@{}l@{}}
  \vote[k] = f_s(L_1, \payload[k]) \sqcap g_s(L_2, \payload[k]);
  \ms
  L_1 = \{ \payload[k'] \mid k' < k \wedge \phase[k'] = \DECIDED
  \wedge {} \\
  \hspace{.8cm}\decision[k'] = \COMMIT\};
  \ms
  L_2 = \{\payload[k'] \mid k' < k \wedge \phase[k'] = \ACCEPTED
  \wedge {}\\
  \hspace{.8cm} \vote[k'] =
  \COMMIT\}.
\end{array}
$$

Then for some $T$, $P$ we have
$$
\begin{array}{@{}l@{}}
  \vote[k] \sqsubseteq f_s(\fploadprocess(T, \txn, \payload),
  \payload[k]) \sqcap {}\\
 \hspace{1.3cm} g_s(\fploadprocess(P, \txn, \payload), \payload[k]) \wedge {}
  \ms
  T = \{ \txn[k'] \mid k' < k \wedge \phase[k'] = \DECIDED \wedge {}\\
  \hspace{.7cm} \decision[k'] = \COMMIT\} \wedge {}
  \ms
  P = \{\txn[k'] \mid k' < k \wedge \phase[k'] = \ACCEPTED \wedge  {}\\
  \hspace{.7cm}\vote[k'] =
  \COMMIT\}.
\end{array}
$$

From the last two conjuncts and Invariant~\ref{inv:committed} we get
$$
\begin{array}{@{}l@{}}
 T = \{ \txn[k'] \mid k' < k \wedge \vote[k'] = \COMMIT \wedge {} \\
\hspace{.7cm} d[\txn[k']] =
\COMMIT\} \setminus P \wedge {}
\ms
(\forall k'.\, \txn[k'] \in T {\implies}\\
\hspace{1.7cm} (\DECISION(\txn[k'], \COMMIT) \mbox{ has been sent})) \wedge {}
\ms
(\forall k' < k.\, \vote[k'] = \COMMIT \wedge \txn[k'] \not\in T \cup P
{\implies} \\
\hspace{1.7cm} (\DECISION(\txn[k'], \ABORT) \mbox{ has been sent})),
\end{array}
$$
which implies the required.

We next consider the transition at line~\ref{alg:accept-vote} by a process
$p_i$. The induction hypothesis implies that, before the
transition at line~\ref{alg:accept-vote}, we have
(\ref{define-process-t-p}) at $p_i$. After processing the
$\ACCEPT(\vballot'', k', t, \pl, d)$, $p_i$ modifies its
$\txn$, $\vote$, $\payload$ and $\phase$ arrays by assigning the $k'$
position. Fix a $k$. We distinguish three cases:
\begin{enumerate}
\item $k<k'$. The required trivially follows from the induction hypothesis.
\item $k=k'$. By Invariant~\ref{inv:synchrony}, after processing the
  $\ACCEPT$ message, $p_i$ has $\prefix{\txn}{k'} \prec \prefix{\vtxn}{k'}$,
  $\prefix{\vote}{k'} \prec \prefix{\vvote}{k'}$ and
  $\prefix{\payload}{k'} \prec \prefix{\vpayload}{k'}$, where $\vtxn$,
  $\vvote$ and $\vpayload$ are the values of the
  arrays $\txn$, $\vote$ and $\payload$ at the leader of $s$ at $\vballot''$ when it sent the
  $\PREPAREACK(\vballot'', s, k', t, \pl, d)$. By the induction hypothesis,
  the leader of $s$ at $\vballot''$ satisfies the
  required before sending the $\PREPAREACK$ message. Hence, by 
the fact that
  $\cf_s$ and $\cg_s$ are distributive, the required is guaranteed at $p_i$ for 
$\vote[k']$ after the transition.
\item $k>k'$. We have that before processing the $\ACCEPT(\vballot'', k', t, \pl, d)$ message,
  $p_i$ has processed $\REPLICATELOGS(\vballot'', \_, \_, \vtxn, \_, \_,
  \_)$. After processing $\REPLICATELOGS$,
  $p_i$ has $\txn=\vtxn$,  $\vote=\vvote$ and $\payload=\vpayload$
  where $\vtxn$, $\vvote$ and $\vpayload$ are the arrays $\txn$,
  $\vote$ and $\payload$ at the leader of $s$ at $\vballot''$ when it
  sent the $\REPLICATELOGS$ message. Let $m=\length(\txn)$ at $p_i$
  after processing $\ACCEPT(\vballot'', k', t, \pl, d)$.

Consider first the case when
  $m=\length(\vtxn)$. Then $p_i$, after processing  the
  $\REPLICATELOGS$ message and before processing $\ACCEPT(\vballot'',
  k', t, \pl, d)$ may have only processed $\ACCEPT(\vballot'', k^*,
  \_, \_, \_)$ such that $k^*\leq m$. Lines~\ref{line:next} and
  ~\ref{line:setnext} trivially guarantee that after processing $\ACCEPT(\vballot'',
  k', t, \pl, d)$, $p_i$ still has $\txn=\vtxn$, $\vote=\vvote$ and
  $\payload=\vpayload$. By the induction hypothesis, the leader of $s$
  at $\vballot''$ satisfies the required before sending the
  $\REPLICATELOGS$. Hence, the required is guaranteed at $p_i$ after
  the transition when $m=\length(\txn)$.

Consider now the case when $m>\length(\txn)$. Therefore, $p_i$ must have received an
  $\ACCEPT(\vballot'', m, \_, \_, \_)$ message and responded to
  it with $\ACCEPTACK$ before processing $\ACCEPT(\vballot'',
  k', t, \pl, d)$. By Invariant~\ref{inv:synchrony}, after processing
  $\ACCEPT(\vballot'', m, \_, \_, \_)$ and while
  $\ballot[s]=\vballot''$, $p_i$ has $\prefix{\txn}{m} \prec \prefix{\vtxn'}{m}$,
  $\prefix{\vote}{m} \prec \prefix{\vvote'}{m}$ and
  $\prefix{\payload}{m} \prec \prefix{\vpayload'}{m}$, where $\vtxn'$,
  $\vvote'$ and $\vpayload'$ are the values of the
  arrays $\txn$, $\vote$ and $\payload$ at the leader of $s$ at $\vballot''$ when it sent the
  $\PREPAREACK(\vballot'', s, m, \_, \_, \_)$. Thus, after processing $\ACCEPT(\vballot'',
  k', \_, \_, \_)$, $p_i$ still has $\prefix{\txn}{m} \prec \prefix{\vtxn'}{m}$,
  $\prefix{\vote}{m} \prec \prefix{\vvote'}{m}$ and
  $\prefix{\payload}{m} \prec \prefix{\vpayload'}{m}$. By the induction hypothesis,
  the leader of $s$ at $\vballot''$ satisfies the
  required before sending $\PREPAREACK(\vballot'', s, m, \_, \_, \_)$. Hence by the fact that
  $\cf_s$ and $\cg_s$ are distributive, the required is guaranteed at $p_i$ after the transition.
\end{enumerate}

Finally, the transitions at lines~\ref{alg:forced-vote} and ~\ref{alg:new-vote}
are handled easily.\qed

\bigskip

We now prove~(\ref{compute1})-(\ref{compute4}). Take the earliest
point in the execution where $d_s[t]$ can be determined as per the
definition given earlier.  Let $\vballot$ be the epoch used in this definition.
Then by Proposition~\ref{prop:vote-compute} at this point, at the
leader of $s$ at $\vballot$ for some $T, P$ we have
\begin{equation}
\nonumber
\begin{array}{@{}l@{}}
d_s[t] \sqsubseteq f_s(\fploadprocess(T,\txn,\payload),
  \fploadprocess(t, \txn,\payload)) \sqcap {}\\
\hspace{1cm}g_s(\fploadprocess(P,\txn,\payload), \fploadprocess(t, \txn,\payload)) \wedge {}
\ms
T = \{\txn[k'] \mid k' < k \wedge \vote[k'] = \COMMIT \wedge {}\\
\hspace{.7cm} d[\txn[k']] = \COMMIT\}
\setminus P_s[t] \wedge {}
\ms
P \subseteq \{\txn[k'] \mid k' < k \wedge \vote[k'] = \COMMIT\}
  \wedge {}
\ms
(\forall k'.\, \txn[k'] \in T {\implies} \\
\hspace{1.7cm}(\DECISION(\txn[k'], \COMMIT) \mbox{ has been sent})) \wedge {}
\ms
(\forall k' < k.\, \vote[k'] = \COMMIT \wedge \txn[k'] \not\in T \cup P
{\implies} \\
\hspace{1.7cm}(\DECISION(\txn[k'], \ABORT) \mbox{ has been sent})).
\end{array}
\end{equation}

For the $T, P$ fixed above, and Invariant~\ref{inv:main} we get
\begin{equation}\label{define-t-p}
\begin{array}{@{}l@{}}
T_s[t] = T =  \{t' \mid \pos_s[t'] < \pos_s[t] \wedge d[t'] = \COMMIT\}
\setminus P_s[t]
\ms
P_s[t] = P\setminus\{t\mid t\in P \wedge\pos_s[t] \mbox{ is not
  defined}\} \subseteq \\
\hspace{1.7cm}\{t' \mid \pos_s[t'] < \pos_s[t] \wedge d_s[t']
  = \COMMIT\}
\ms
(\forall k'.\, \txn[k'] \in T_s[t]{\implies} \\
\hspace{1.7cm} (\DECISION(\txn[k'], \COMMIT) \mbox{ has been sent})) \wedge {}
\ms
(\forall k' < k.\, \vote[k'] = \COMMIT \wedge \txn[k'] \not\in T_s[t] \cup P_s[t]
{\implies}\\
\hspace{1.7cm} (\DECISION(\txn[k'], \ABORT) \mbox{ has been sent})),
\end{array}
\end{equation}
which establishes~(\ref{compute2}) and (\ref{compute4}).

By Invariant~\ref{inv:payload},
(\ref{property-payload}) and by the fact that $\cf_s$ and $\cg_s$ are
distributive, we get
$$
d_s[t] \sqsubseteq f_s(\pload_s (T_s[t]), \pload_s[t]) \sqcap
  g_s(\pload_s (P_s[t]), \pload_s[t]),
$$
which establishes~(\ref{compute1}) for the $T_s[t], P_s[t]$
fixed above.

Finally, we prove~(\ref{acyclic}). To this end, we show that if $t'\rt t$ or
$t' \decrel t$, then a $\DECISION(t', d[t'])$ message was sent in the execution,
and this had happened before any $\DECISION(t, \_)$ message was sent.  The case
of $t'\rt t$ is trivial and therefore we only consider the case of
$t'\decrel t$.  Take the earliest point in the execution where we can define
$d_s[t]$, and hence, $T_s[t]$ and $P_s[t]$ (by~(\ref{define-t-p})). Then a
$\DECISION(t, \_)$ message could not have been sent by this point. Assume first
that $t' \in T_s[t]$. Then by~(\ref{define-t-p}) a $\DECISION(t', \COMMIT)$
message has been sent earlier. Now assume that
$$
\pos_s[t'] < \pos_s[t] \wedge d_s[t'] = \COMMIT \wedge d[t'] = \ABORT \wedge t'
\not\in P_s[t].
$$
Then at this point $\txn[\pos_s[t']] = t'$ and $\vote[\pos_s[t']] =
\COMMIT$, so that by~(\ref{define-t-p}) a
$\DECISION(t', \ABORT)$ message has been sent earlier. We have thus proved (\ref{acyclic}).\qed

\begin{lemma}\label{thm:2pc}
  If shard-local certification functions $f_s$ and $g_s$
  satisfy~(\ref{local-global})-(\ref{g-conflict-f}), then
  every history satisfying \lltcsone{} is correct with respect
  to $f$.
\end{lemma}

\subparagraph*{Proof.} This follows the proof of Theorem~1
in~\cite{discpaper}\footnote{Appendix A of {\tt
    https://arxiv.org/pdf/1808.00688}.}  with minimal adjustments.\qed

\subparagraph*{Proof of Theorem~\ref{thm:conc}.} Follows from
Lemmas~\ref{thm:constraints} and~\ref{thm:2pc}.\qed

\section{Proof of Liveness}
\label{sec:proof-liveness}

We prove the nontrivial Theorem~\ref{thm:livenessreconfiguration}:

\begin{quote}
  If a process $p_r$ attempts to reconfigure a shard $s$ and no other process
  attempts to reconfigure $s$ simultaneously, then if $p_r$ is non-faulty for long
  enough, it will eventually introduce a new configuration.
\end{quote}

\subparagraph*{Proof.} Assume that a process $p_r$ attempts to
reconfigure a shard $s$. Take the earliest point in the execution
where $p_r$ calls the {\tt reconfigure} function. Let $\vballot$ be
the epoch of the last active configuration of $s$ at that point in time. The
process $p_r$ first queries the \external to find the latest
introduced configuration
of $s$ to start the probing.
Let $\vballot'$ be the epoch of this configuration.

Assume that the probing eventually ends. After this happens, $p_r$ computes the membership of the new
configuration $c$ (lines~\ref{line:members}).
Then $p_r$ attempts to write $c$ into the \external. Since there is not
other process attempting to reconfigure $s$ simultaneously, $p_r$ will
succeed. This last step introduces $c$, as required.

We now prove that the probing eventually ends, provided that no other process
attempts reconfiguring $s$ simultaneously and $p_r$ is non-faulty for
long enough. The probing procedure proceeds by iterations in
epoch descending order, starting by probing the members of
$s$ at $\vballot'$. The process $p_r$ only moves to the next iteration after receiving at least one reply from a member of $s$ at the epoch being
currently probed while no process replies with $\PROBEACK(\TRUE,
\vballot+1, s)$. Consider an arbitrary epoch $\vballot''$ such that $\vballot''\leq\vballot'$. If $p_r$ is probing the
members of $s$ at $\vballot''$, then $p_r$ has received a
$\PROBEACK(\FALSE, \vballot+1, s)$ from at least one member of $s$ at
each epoch $\vballot^*$ such that
$\vballot''<\vballot^*\leq\vballot'$. Furthermore, because of
line~\ref{line:probe-set-newepoch} and the check in line~\ref{alg:check-replicatelogs}, none of these configurations will ever become
active. Then by Assumption~\ref{liveness-condition} and 
the fact that there is no concurrent reconfiguration, $p_r$
is guaranteed to receive at least one reply from a member of $s$ at
$\vballot''$. Hence, for each epoch $\vballot''$ that $p_r$ probes, either the
whole probing terminates, or $p_r$ will eventually move to probe the previous
epoch. Assume that $p_r$ reaches epoch $\vballot$. By
line~\ref{line:probe-set-newepoch} and the check in line~\ref{alg:check-replicatelogs}, the configurations of $s$ with epoch
$\vballot^*$, such that $\vballot<\vballot^*\leq\vballot'$, will never become
active. Then by Assumption~\ref{liveness-condition} and 
the fact that there is no concurrent reconfiguration, $p_r$
is guaranteed to receive at least one reply from a member of $s$ at
$\vballot$. That the configuration of $s$ with epoch $\vballot$ became
active implies that every member of $s$ at $\vballot$ has $\recovery=\TRUE$ when being probed. Hence, the
probing procedure is guaranteed to finish.\qed

\section{RDMA-based Atomic Commit Protocol}
\label{sec:rdma}

We give the pseudocode of the RDMA-based protocol in Figures~\ref{fig:normal-rdma}
and~\ref{fig:reconf-rdma}. The redesigned reconfiguration protocol
uses a slightly different set of variables. Instead of the variable $\probing$, the
protocol uses the variable $\reconstatus\in\{\READYSTATUS, \PROBING,
\INSTALLING\}$ to record whether a process is \emph{ready} to start
reconfiguring the system, \emph{probing} the system or
\emph{disseminating} a new configuration. A variable
$\openconnections\in 2^{\Proc}$ records the set of processes to which a process
currently maintains
an open RDMA connection. Also, the variables $\pballot$ and
$\pmembers$ are now arrays:
$\pballot\in \Shard\to \mathbb{N}$ and $\pmembers\in \Shard\to
2^{\Proc}$. This change is required because now reconfiguration
involves all shards, instead of a single one. Finally, the
data structures maintained by the external \external and its interface are
adjusted as well. Instead of keeping a separate
data structure with each shard's sequence of configurations, the \external
keeps a single data structure with the system's sequence of configurations
parameterized by shard. Moreover, none of the three operations of the
\external's interface take a shard identifier as argument anymore.

To prove the correctness of the protocol,
apart from the set of invariants (Figures~\ref{fig:inv} and~\ref{fig:inv-rest}) used to prove the correctness of
the atomic commit protocol in Figure~\ref{fig:protocol}, we require
the following invariant, formalizing property (*) from \S\ref{sec:rdmadiscussion}:
\begin{enumerate}[leftmargin=15pt, label={\arabic*.}, ref={\arabic*}]
\setcounter{enumi}{12}
\item \label{inv-rdma:epoch} Assume that the coordinator of a transaction $t$ receives a
$\PREPAREACK(\vballot, s, k, t, \pl, d)$ message and sends an
$\ACCEPT(k, t, \pl, d)$ message to a process $p_i$. If $p_i$
receives the $\ACCEPT(k, t, \pl, d)$ message, then it has
$\ballot=\vballot$ right before this.
\end{enumerate}

This invariant trivially holds in
the atomic commit protocol in Figure~\ref{fig:protocol} but
is nontrivial in the RDMA-based protocol.

We first prove Invariant~\ref{inv-rdma:epoch}. Then we prove
Invariants~\ref{inv:probe} and~\ref{inv:synchrony}, whose proofs
rely on Invariant~\ref{inv-rdma:epoch}. We skip the
proofs for the rest of the invariants, as these
are similar to the proofs of the invariants in Figures~\ref{fig:inv}
and~\ref{fig:inv-rest} of the protocol in Figure~\ref{fig:protocol}, with small adjustments due to differences in
the protocols' pseudocodes. Finally, we prove
the following theorem.

\begin{theorem}\label{thm-rdma:conc}
  A transaction certification service implemented using the protocol in
  Figures~\ref{fig:normal-rdma} and~\ref{fig:reconf-rdma} is correct with respect
  to a certification function $\cf$ matching the shard-local certification
  functions $\cf_s$ and $\cg_s$.
\end{theorem}


\subparagraph*{Proof of Invariant~\ref{inv-rdma:epoch}.}  Assume that the
coordinator $p_c$ of a transaction $t$ receives a
$\PREPAREACK(\vballot, s, k, t, \pl, d)$ message and sends an
$\ACCEPT(k, t, \pl, d)$ message to a process $p_i$. Assume further that $p_i$
receives the $\ACCEPT(k, t, \pl, d)$ message and let $\ballot=\vballot'$ at
$p_i$ right before this transition. We prove that $\vballot'=\vballot$.

The leader $p_l$ of $s$ at $\vballot$ must have received
$\PREPARE(t, \pl)$ and replied with
$\PREPAREACK(\vballot, s, k, t, \pl, d)$, and when it received the message, it
had $\ballot=\vballot$. Thus, $p_l$ must have received $\NEWCONFIG(\vballot)$
earlier. Also, by the check in line~\ref{alg-rdma:check-prepareack}, $p_i$ must
be a member of $\vballot$. Then $p_i$ had processed
$\CONFIGPREPARE(\vballot, \_, \_)$ before $p_c$ sent the $\ACCEPT(k, t, \pl, d)$
message to $p_i$. When $p_i$ processed $\CONFIGPREPARE(\vballot,\_, \_)$, it had
no open connections, either because it was probed
(line~\ref{alg-rdma:probe-close}) or because it is a new process. The process
$p_i$ only opens them, allowing it to receive $\ACCEPT(k, t, \pl, d)$, when it
receives either a $\NEWCONFIG(\vballot^*)$ or a
$\REPLICATELOGS(\vballot^*,\_, \_,\_,\_,\_)$ message, so that
$e' \ge \vballot^*$. By the fact that $\cballot$ gets updated when processing
$\CONFIGPREPARE$ and by the checks in lines~\ref{alg-rdma:check-newconfig}
and~\ref{alg-rdma:check-newstate}, we have that $\vballot^*\geq\vballot$. Then
$\vballot' \ge \vballot$.

Assume now that $\vballot'>\vballot$. When $p_i$ processes $\ACCEPT(k, t, \pl,
d)$ it has $\ballot=\vballot'$. Then $p_i$ has received $\NEWCONFIG(\vballot')$ or $\REPLICATELOGS(\vballot',
\_,\_,\_,\_,\_)$ before. When processing any of these messages, $p_i$
has no open connections. Therefore, $p_c$ must have sent $\ACCEPT(k, t, \pl,
d)$ after $p_i$ processed $\NEWCONFIG(\vballot')$ or $\REPLICATELOGS(\vballot',
\_,\_,\_,\_,\_)$. Furthermore, by the checks in
lines~\ref{alg-rdma:receive-connect}
and~\ref{alg-rdma:receive-connect-ack}, $p_i$ only establishes
connections to the
members of $\vballot'$. Thus, $p_c$ must be
a member of $\vballot'$ to successfully send $\ACCEPT(k, t, \pl,
d)$ to $p_i$. Then $p_c$ must have
received $\CONFIGPREPARE(\vballot', \_, \_)$ and replied with
$\CONFIGPREPAREACK(\vballot')$ before $p_i$ processed $\NEWCONFIG(\vballot')$ or $\REPLICATELOGS(\vballot',
\_,\_,\_,\_,\_)$ and therefore before sending $\ACCEPT(k, t, \pl,
d)$. When processing $\CONFIGPREPARE(\vballot', \_, \_)$, $p_c$ has no
open connections. Thus, $p_c$ can only send $\ACCEPT(\vballot, k, t, \pl,
d)$ after receiving $\NEWCONFIG(\vballot^*)$ or $\REPLICATELOGS(\vballot^*,
\_,\_,\_,\_,\_)$, where
$\vballot^*\geq\vballot'$. This implies that $p_c$ received
$\PREPAREACK(\vballot, s, k, t, \pl, d)$ after setting
$\ballot=\vballot^* \ge e' > e$. By the
check in line~\ref{alg-rdma:check-prepareack}, $p_c$ then would never send $\ACCEPT(k, t, \pl,
d)$ to $p_i$ at this point. Hence, we must have $\vballot' \le \vballot$, which
together with $\vballot' \ge \vballot$ implies $\vballot' = \vballot$.\qed

\subparagraph*{Proof of Invariant~\ref{inv:synchrony}.} 
Assume that a
process $p_i$ in $s$ at $\vballot$ processes $\ACCEPT(k, t, \pl,
d)$. We prove that, after the transition
and while $\ballot[s]=\vballot$, $p_i$ has $\prefix{\txn}{k} \prec
\prefix{\vtxn}{k}$, $\prefix{\vote}{k} \prec
\prefix{\vvote}{k}$ and $\prefix{\payload}{k} \prec
\prefix{\vpayload}{k}$, where $\vtxn$, $\vvote$ and $\vpayload$ are the values of the
arrays $\txn$, $\vote$ and $\payload$ at the leader $p_l$ of $s$ at $\vballot$ when it
sent the corresponding message $\PREPAREACK(\vballot, s, k, t, \pl, d)$. 

By Invariant~\ref{inv-rdma:epoch}, if $p_i$ processes $\ACCEPT(k, t, \pl,
d)$, then $p_i$ has $\ballot=\vballot$. Thus, $p_i$ has processed $\REPLICATELOGS(\vballot,
\vtxn',\vpayload',\vvote',\_,\_)$ before. After processing this message, $p_i$ has $\txn=\vtxn'$,
$\vote=\vvote'$ and $\payload=\vpayload'$ where $\vtxn'$, $\vvote'$ and $\vpayload'$ are the values of the
arrays $\txn$, $\vote$ and $\payload$ at the leader $p_l$ of $s$ at
$\vballot$ when it sent the $\REPLICATELOGS$ message. Let
$k'=\length(\vtxn')$. By lines~\ref{line-rdma:unique_accept},~\ref{alg-rdma:next} and ~\ref{alg-rdma:setnext} we have that $\prefix{\vtxn}{k'} =
\vtxn'$, $\prefix{\vvote}{k'} = \vvote'$ and $\prefix{\vpayload}{k'} =
\vpayload'$. By Invariant~\ref{inv-rdma:epoch}, $p_i$ has $\prefix{\txn}{k'} \prec \prefix{\vtxn}{k'}$,
$\prefix{\vote}{k'} \prec \prefix{\vvote}{k'}$ and
$\prefix{\payload}{k'} \prec \prefix{\vpayload}{k'}$ while
$\ballot[s]=\vballot$. Furthermore, after processing $\ACCEPT(k, t, \pl,
d)$, $p_i$ has $\txn[k]=\vtxn[k]$, $\vote[k]=\vvote[k]$ and
$\payload[k]=\vpayload[k]$. 
By Invariants~\ref{inv:accept-tx} and~\ref{inv-rdma:epoch}, $p_i$ has $\txn[k]=\vtxn[k]$, $\vote[k]=\vvote[k]$ and
$\payload[k]=\vpayload[k]$ while $\ballot[s]=\vballot$. 

We now prove that after processing $\REPLICATELOGS(\vballot,
\vtxn',\vpayload',\vvote',\_,\_)$ and while
$\ballot[s]=\vballot$, $p_i$ has $\txn[k'']\in\{\vtxn[k''], \bot\}$,
$\vote[k'']\in\{\vvote[k''], \bot\}$ and
$\payload[k'']\in\{\vpayload[k''], \bot\}$ for any $k''$ such that
$k'<k''<k$. We prove it by induction on
the length of the protocol execution from the moment in which $p_i$ has
processed $\REPLICATELOGS(\vballot,
\vtxn',\vpayload',\vvote',\_,\_)$. The validity of the
property can be affected by only the transition at
line~\ref{alg-rdma:receive-accept}. Let $\ACCEPT(k^*, t^*, \pl^*,
d^*)$ be the message that triggers the transition. Assume that
$k'<k^*<k$, as otherwise the transition does not affect the validity
of the property. By the induction hypothesis, $p_i$ has $\txn[k'']\in\{\vtxn[k''], \bot\}$,
$\vote[k'']\in\{\vvote[k''], \bot\}$ and
$\payload[k'']\in\{\vpayload[k''], \bot\}$ for any $k''\neq
k^*$ such that $k'<k''<k$ after processing $\ACCEPT(\vballot, k^*, t^*, \pl^*,
d^*)$. Also, after processing $\ACCEPT(\vballot, k^*, t^*, \pl^*,
d^*)$, $p_i$ has $\txn[k^*]=t^*$, $\vote[k^*]=d^*$ and
$\payload[k^*]=\pl^*$. By
Invariant~\ref{inv-rdma:epoch}, $t^*$ must have been prepared by the leader $p_l$ of
$s$ at $\vballot$. Then, by lines~\ref{line-rdma:unique_accept},~\ref{alg-rdma:next} and ~\ref{alg-rdma:setnext}, $\vtxn[k^*]=t^*$, $\vvote[k^*]=d^*$ and
$\vpayload[k^*]=\pl^*$. Then, after processing $\ACCEPT(\vballot, k^*, t^*, \pl^*,
d^*)$, $p_i$ has $\txn[k^*]=\vtxn[k^*]$, $\vote[k^*]=\vvote[k^*]$ and
$\payload[k^*]=\vpayload[k^*]$. We have already proved that (i) $p_i$ has $\prefix{\txn}{k'} \prec \prefix{\vtxn}{k'}$,
$\prefix{\vote}{k'} \prec \prefix{\vvote}{k'}$ and $\prefix{\payload}{k'}
\prec \prefix{\vpayload}{k'}$ after processing $\REPLICATELOGS(\vballot,
\vtxn',\vpayload',\vvote',\_,\_)$ and while
$\ballot[s]=\vballot$; and that (ii) $p_i$ has $\txn[k]=\vtxn[k]$, $\vote[k]=\vvote[k]$ and
$\payload[k]=\vpayload[k]$ after processing $\ACCEPT(\vballot, k, t, \pl,
d)$ and while $\ballot[s]=\vballot$. Hence, $p_i$ has $\prefix{\txn}{k} \prec
\prefix{\vtxn}{k}$, $\prefix{\vote}{k} \prec
\prefix{\vvote}{k}$ and $\prefix{\payload}{k} \prec
\prefix{\vpayload}{k}$ after processing $\ACCEPT(\vballot, k, t, \pl,
d)$ and while $\ballot[s]=\vballot$, as required. \qed

\subparagraph*{Proof of Invariant~\ref{inv:probe}.} When $p_i$ processed $\PROBE(\vballot)$, it
closed all connections. This prevents $p_i$ from acknowledging any $\ACCEPT$ message
until it processes a $\NEWCONFIG(\vballot^*)$ or a
$\REPLICATELOGS(\vballot^*, \_,\_,\_,\_,\_)$. When $p_i$
processed $\PROBE(\vballot)$, it also sets
$\cballot=\vballot$. By the checks in lines~\ref{alg-rdma:check-newconfig}
  and~\ref{alg-rdma:check-newstate} and by the fact that $\cballot$
  can never decrease, this guarantees that $p_i$ only
handles any of these messages if $\vballot^*\geq e$. Hence,
by the time $p_i$ is able to process $\ACCEPT$ messages again it will have
$\ballot=\vballot^*>\vballot'$. Since $\ballot$ never decreases at a process, 
from this point on, by Invariant~\ref{inv-rdma:epoch},
$p_i$ will not process any $\ACCEPT$ message prepared in an epoch
preceding $\vballot^*$, as required.\qed

\begin{lemma}\label{thm-rdma:constraints}
  The atomic commit protocol in
  Figures~\ref{fig:normal-rdma} and~\ref{fig:reconf-rdma} is a correct
  implementation of \lltcsone{}.
\end{lemma}

\subparagraph*{Proof} This follows the proof of Lemma~\ref{thm:constraints} with
minimal adjustments.\qed

\subparagraph*{Proof of Theorem~\ref{thm-rdma:conc}.} Follows from
Lemmas~\ref{thm-rdma:constraints} and~\ref{thm:2pc}.\qed

\begin{figure*}[p]
\begin{tabular}{@{}l@{}|@{\quad\quad}l@{}}
\begin{minipage}[t]{8.8cm}
\removelatexerror
\begin{algorithm}[H]
  \SubAlgo{\Fun \Certify($t, \pl$)}{
    \ForAll{$s\in \shards(t)$}{
    \Send $\PREPARE(t, (\pl\mid s))$ \KwTo $\leader[s]$;
    }
  }

  \smallskip
  \smallskip
  \smallskip

  \SubAlgo{\WhenReceived $\PREPARE(t, \pl)$ {\bf from
      $p_j$}}{
     \textbf{pre:} $\status = \LEADER$\;
    \uIf{$\exists k.\, t = \txn[k]$}{\label{line-rdma:unique_accept}
        \Send $\PREPAREACK(\ballot, s_0, k, \txn[k],$\\
        \nonl \hspace{2.58cm}$\payload[k], \vote[k])$ \KwTo $p_j$}
    \Else{
      $\nextv \leftarrow \nextv + 1$\;\label{alg-rdma:next}
      $(\txn,\phase)[\nextv] \leftarrow  (t, \ACCEPTED)$\;
      \uIf{$\pl\neq \bot$}{
      $\vote[\nextv] \leftarrow f_{s_0}(\setpayload_1, \pl) \sqcap
      g_{s_0}(\setpayload_2,\pl)$\;\label{alg:compute-vote-rdma} 
      $\payload[\nextv] \leftarrow \pl$\;
      }\Else{
         $\vote[\nextv] \leftarrow
         \ABORT$\;\label{alg-rdma:forced-vote}
         $\payload[\nextv] \leftarrow \varepsilon$\;
      }
      \Send $\PREPAREACK(\ballot, s_0, \nextv, t, $\\
        \nonl \hspace{2.58cm}$\payload[\nextv], \vote[\nextv])$ \KwTo $p_j$;
    }
  }

\end{algorithm}
\end{minipage}
&
\begin{minipage}[t]{8.8cm}
\removelatexerror
\begin{algorithm}[H]
\SubAlgo{\WhenReceived $\PREPAREACK(\vballot, s, k, t, \pl,
    d)$}{
    \textbf{pre:} $\vballot = \ballot$\; \label{alg-rdma:check-prepareack}
    \sendrdma $\ACCEPT(k, t, \pl, d)$ \KwTo $\M[s]\setminus \leader[s]$;
    }

\smallskip
  \smallskip
  \smallskip

  \SubAlgo{\WhenDelivered $\ACCEPT(k, t, \pl, d)$ {\bf from $p_j$}}{\label{alg-rdma:receive-accept}
      $(\txn, \payload, \vote, \phase)[k] \leftarrow (t, \pl, d, \ACCEPTED)$\; \label{alg-rdma:accept-vote}    
  }

  \smallskip
  \smallskip
  \smallskip

  \SubAlgo{{\bf when for every $s \in \shards(t)$ and
      $p_j\in \M[s]\setminus \leader[s]$ received an \ackrdma from $p_j$ for
      $\ACCEPT(k_s, t, \pl_s, d_s)$ sent in response to
      $\PREPAREACK(\vballot, s, k_s, t, \pl_s,
      d_s)$}}{\label{alg-rdma:receive-acceptack} \textbf{pre:}
    $\vballot = \ballot$\; \Send
    $\DECISION(t, \bigsqcap_{s \in \shards(t)} d_s)$ {\bf to $\client(t)$}\;
    \ForAll{$s \in \shards(t)$}{
      \sendrdma $\DECISION(k_s, \bigsqcap_{s \in \shards(t)} d_s)$\\
      \nonl \quad \KwTo $\M[s]$;} }

  \smallskip
  \smallskip
  \smallskip

  \SubAlgo{\WhenDelivered $\DECISION(k, d)$}{\label{alg-rdma:receive-decision}
    $(\decision, \phase)[k] \leftarrow (d, \DECIDED)$\;\label{alg-rdma:decision-store}
      }

\end{algorithm}
\end{minipage}
\end{tabular}
\\
\caption[center]{\rm RDMA-based protocol at a process $p_i$ in a shard $s_0$:
  failure-free case. At line~\ref{alg:compute-vote-rdma} we let\\
  $\setpayload_1=\{\payload[k] \mid k < \nextv \wedge\phase[k] = {\normalfont \DECIDED} \wedge
\decision[k] = {\normalfont \COMMIT}\}$;\\
$\setpayload_2=\{\payload[k] \mid k < \nextv \wedge \phase[k] =
{\normalfont \ACCEPTED} \wedge
\vote[k] = {\normalfont \COMMIT}\}$.}
\label{fig:normal-rdma}
\end{figure*}

\begin{figure*}[p]
\begin{tabular}{@{}l@{}|@{\quad\quad}l@{}}
\begin{minipage}{9.2cm}
\removelatexerror
\begin{algorithm}[H]

  \SubAlgo{\Fun ${\tt reconfigure}()$}{ 
    \textbf{pre:} $\reconstatus=\READYSTATUS$\;
    $\reconstatus\gets \PROBING$\;
    $\langle \vballot, \pmembers, \_ \rangle \gets$ {\tt
      get\_last}() {\bf at} $\zk$\;
    \ForAll{$s\in\Shards$}{
    $\pballot[s]\gets \vballot$\;}
    $\rballot\gets \vballot+1$\;
    \Send $\PROBE(\rballot)$ \KwTo $\pmembers$\;
    }

    \smallskip
    \smallskip
    \smallskip
    
 \SubAlgo{\WhenReceived $\PROBE(\vballot)$ \KwFrom
    $p_j$}{
    \textbf{pre:} $\vballot \geq\cballot$\;
    $\status\gets\RECOVERING$\;
     \multiclose($\openconnections$)\;\label{alg-rdma:probe-close}
    $\cballot\gets \vballot$\;
    \Send $\PROBEACK(\recovery, \vballot, s_0)$ \KwTo $p_j$\;
    }

  \smallskip
    \smallskip
    \smallskip
  
\SubAlgo{{\bf when for every $s \in \shards(t)$ received a}
      $\PROBEACK(\TRUE, \rballot, s)$}{
      \textbf{pre:} $\reconstatus=\PROBING$\;
      $\reconstatus\gets \READYSTATUS$\;
     $\langle\rmembers, \rleaders\rangle\gets$ {\tt compute\_membership}()\;
    {\bf var} $r\gets$ {\tt compare\_and\_swap} ($\rballot-1,$\\
    \nonl\hspace{0.15cm}$\langle \rballot, \rmembers, \rleaders \rangle$) {\bf at} $\zk$\;
    \If{$r$}{
           $\reconstatus\gets \INSTALLING$\;

        \Send $\CONFIGPREPARE (\rballot, \rmembers,$\\
         \nonl \hspace{0.25cm}$ \rleaders)$ \KwTo $\rmembers$\;

  }
}

\smallskip
    \smallskip
    \smallskip

  \SubAlgo{{\bf non-deterministically when received} $\PROBEACK(\FALSE,
  \rballot, s)$ {\bf from}
  $p_j \in \pmembers[s]$ {\bf and no} $\PROBEACK(\TRUE, \rballot, s)$}{
  \textbf{pre:} $\reconstatus=\PROBING$\;
  $\pballot[s]\gets\pballot[s]-1$\;
  $\langle\_, \vm, \_ \rangle \gets$ {\tt get}($\pballot[s]$) {\bf
    at} $\zk$\;
  $\pmembers[s]\gets \vm[s]$\;
   \Send $\PROBE(\rballot)$ \KwTo $\pmembers[s]$\;
}

  \smallskip
    \smallskip
    \smallskip

\SubAlgo{\WhenReceived $\CONFIGPREPARE(\vballot, \vm, leaders)$ \KwFrom $p_j$}{
    \textbf{pre:} $\vballot\geq\cballot $\;
    $\M \gets \vm$\;
    $\leader \gets leaders$\;
      $\cballot=\vballot$\;
    \Send $\CONFIGPREPAREACK(\vballot)$ \KwTo $p_j$\;
}

\end{algorithm}
\end{minipage}
&
\begin{minipage}{9cm}
\removelatexerror
\begin{algorithm}[H]  

\SubAlgo{\WhenReceived $\CONFIGPREPAREACK(\vballot)$ \textbf{from every} $p_j\in
      \rmembers$}{\label{alg:receive-configprepareack}
      \textbf{pre:} $\reconstatus=\INSTALLING$\;
    \Send $\NEWCONFIG(\vballot)$ \KwTo $\rleaders$\;
    $\reconstatus\gets \READYSTATUS$\;
}

\smallskip
    \smallskip
    \smallskip

\SubAlgo{\WhenReceived $\NEWCONFIG(\cballot)$ \KwFrom
    $p_j$}{\label{alg-rdma:check-newconfig}
    \emptybuffers()\;
    $\status\gets\LEADER$\;
    $\ballot\gets \cballot$\;
    $\nextv\gets \max\{k \mid \phase[k] \not=
    \START\}$\; \label{alg-rdma:setnext}
    \Send $\REPLICATELOGS(\ballot, \txn, \payload,$\\
    \nonl \hspace{2.25cm}$ \vote, \dec, \phase)$ \KwTo
    $\M[s_0]\setminus p_i$\;
     \Send $\OPENCONNECTION(\ballot)$ \KwTo $\M\setminus \{p_i\}$\;
}

  \smallskip
    \smallskip
    \smallskip

\SubAlgo{\WhenReceived $\REPLICATELOGS(\cballot, \vtxn, \vpayload, $\\ 
    \nonl \hspace{3.55cm} $\vvote,
    \vdecision, \vphase)$ \KwFrom
    $p_j$}{\label{alg-rdma:check-newstate}
    $\status\gets\FOLLOWER$\;
    $\ballot\gets \cballot$\;
    $\recovery\gets \TRUE$\;
    $(\txn, \payload, \vote, \decision, \phase)  \leftarrow (\vtxn,
    \vpayload, \vvote, \vdecision, \vphase)$\; \label{alg-rdma:new-vote}
     \Send $\OPENCONNECTION(\ballot)$ \KwTo $\M\setminus \M[s_0]$\;
    }

\smallskip
    \smallskip
    \smallskip

\SubAlgo{\WhenReceived $\OPENCONNECTION(\ballot)$ \KwFrom
    $p_j$}{\label{alg-rdma:receive-connect}
   \textbf{pre:} $\status\neq\RECOVERING\wedge p_j\not\in\openconnections$\;
   \openconnection($p_j$)\;
   $\openconnections\gets\openconnections \cup \{p_j\}$\;
 \Send $\OPENCONNECTIONACK(\ballot)$ \KwTo $p_j$\;
}

\smallskip
    \smallskip
    \smallskip

\SubAlgo{\WhenReceived $\OPENCONNECTIONACK(\ballot)$ \KwFrom
    $p_j$}{\label{alg-rdma:receive-connect-ack}
   \textbf{pre:} $\status\neq\RECOVERING\wedge p_j\not\in\openconnections$\;
   \openconnection($p_j$)\;
   $\openconnections\gets\openconnections \cup \{p_j\}$;
}

\smallskip
    \smallskip
    \smallskip

\SubAlgo{\Fun \multiclose($P$)}{
    \ForAll{$p\in P$}{
    \closeconnection($p$)\;
    $\openconnections\gets\openconnections \setminus \{p\}$\;
  }}

\smallskip
    \smallskip
    \smallskip

  \SubAlgo{\Fun ${\tt retry}(k)$}{
    \textbf{pre:} $\phase[k] = \ACCEPTED$\;
    \ForAll{$s\in \shards(\txn[k])$}{
    \Send $\PREPARE(\txn[k], \bot)$ \KwTo $\leader[s]$;
  }}

  \smallskip
    \smallskip
    \smallskip

\end{algorithm}
\end{minipage}
\end{tabular}
\\
\caption{\rm RDMA-based protocol at a process $p_i$ in a shard $s_0$:
  reconfiguration.}
\label{fig:reconf-rdma}
\end{figure*}



\fi

\end{document}